# Chaos and Noise in Galactic Potentials

**Salman Habib***, **Henry E. Kandrup**†‡, **and M. Elaine Mahon**†

**T-6, Theoretical Astrophysics*
*and*
*T-8, Elementary Particles and Field Theory*
*Los Alamos National Laboratory*
*Los Alamos, New Mexico 87545*

*†Department of Astronomy and Institute for Fundamental Theory*
*and*
*‡Department of Physics*
*University of Florida*
*Gainesville, Florida 32611*

e-mail:
habib@eagle.lanl.gov
kandrup@astro.ufl.edu
mahon@astro.ufl.edu

**Abstract**

This paper summarises an investigation of the effects of weak friction and noise in time-independent, nonintegrable potentials which admit both regular and stochastic orbits. The aim is to understand the qualitative effects of internal and external irregularities associated, *e.g.*, with discreteness effects or couplings to an external environment, which stars in any real galaxy must experience. One principal conclusion is that these irregularities can be important already on time scales much shorter than the natural relaxation time scale $t_R$ associated with the friction and noise. When viewed in terms of their effects on the collisionless invariants, friction and noise serve simply to induce a classical diffusion process, significant changes in quantities like the energy arising only on a time scale $t_R$. However, when viewed in configuration or velocity space, they have more complicated effects which depend on whether the orbits are regular or stochastic. In particular, for stochastic orbits friction and noise induce an average exponential divergence from the unperturbed Hamiltonian trajectory at a rate which is set by the value of the local Lyapunov exponent. This implies that even weak friction and noise can make a *pointwise* interpretation of orbits suspect already on very short time scales $\ll t_R$. If the phase space contains large measures of both regular and stochastic orbits, the friction and noise can also have significant effects on the *statistical* properties of ensembles of stochastic orbits. Islands of regularity embedded in the stochastic sea are typically surrounded by cantori, which divide stochastic orbits into two classes: confined or sticky stochastic orbits which are trapped near an island of regularity, and filling stochastic orbits which travel unimpeded throughout the stochastic sea. The cantori do not serve as an absolute barrier, so that transitions between confined and filling stochastic orbits are possible. In the absence of friction and noise, this is a very slow diffusion process, proceeding only on a time scale $t \gg 100t_{cr}$, with $t_{cr}$ a typical crossing time. Over short times, it thus makes sense to identify two separate classes of stochastic orbits, each characterised by its own near-invariant measure. However, even very weak friction and noise can drastically accelerate this diffusion. In certain cases, friction and noise corresponding to $t_R \geq 10^6 t_{cr}$ can induce transitions for more than half the orbits within a Hubble time, $t_H \sim 100t_{cr}$. Thus, the distinction between confined and filling stochastic orbits becomes blurred, so that any notion of a noisy near-invariant measure must allow for a mixture of both confined and filling orbits. Potential implications for galactic dynamics are discussed, including the problem of shadowing.

*Subject headings:* galaxies: evolution – galaxies: kinematics and dynamics – galaxies: structure

# 1 Introduction and Motivation

The theoretical foundations of galactic dynamics go back over fifty years to a time when most physicists and astronomers had a view of the physical world that was dominated by integrable or near-integrable systems. The principal focus was on the identification of collisionless invariants appropriate for isolated systems in a true equilibrium. However, over the past several decades, it has been recognised that chaotic phenomena can play an important role in various systems and, as discussed in Section 2, there is also evidence that chaos may be important in galactic dynamics as well. It is therefore useful, and important, to reexamine some of the basic assumptions of conventional galactic dynamics in the context of nonintegrable mean field potentials which admit both regular and stochastic orbits.

In recent years, a good deal of attention has focused on the construction of self-consistent galactic models, idealised as time-independent solutions to the collisionless Boltzmann equation. Given various simplifying assumptions about the shape of the equilibrium, it is possible to construct exact analytic solutions, such as the integrable Stäckel models recently popularised by de Zeeuw (1985). In general, however, analytic methods are inadequate, and one is forced to employ numerical techniques to construct approximate equilibria (cf. Schwarzschild 1979).

Because these techniques aim to construct equilibrium solutions to the collisionless Boltzmann equation, they have involved a consideration of ensembles of Hamiltonian orbits in smooth time-independent potentials. This approach therefore completely ignores discreteness effects, *i.e.*, collisionality, and other internal irregularities, as well as influences resulting from a coupling to an external environment. However, in the past several decades it has become increasingly evident that galaxies cannot always be idealised as completely isolated entities (cf. Toomre & Toomre 1972, Schweizer 1986, Zepf & Whitmore 1993); and, as will be discussed below (cf. Pfenniger 1986), there is also reason to believe that, in certain circumstances, discreteness effects may be more important than is generally recognised. It is therefore important to investigate the possible implications of external perturbations and various internal irregularities on approximate self-consistent equilibria.

A detailed rigourous analysis of these effects is necessarily quite complicated, and oftentimes impossible: In order to allow correctly for all discreteness effects, one must really consider the full $N$-body problem, as formulated in the $N$-particle phase space. A proper treatment of external influences requires a detailed knowledge of the environment, which may not be available observationally. However, there is solid reason to believe that, in many cases, these internal and external effects can be modelled phenomenologically in terms of friction and noise, the forms of which are related by a fluctuation-dissipation theorem (cf. Kubo *et al* 1991).

The idea that discreteness effects, *i.e.*, collisionality, can be modelled as friction and noise lies at the heart of collisional stellar dynamics, as formulated originally by Chandrasekhar (1942, 1943c). From the viewpoint of the nonlinear dynamicist, the justification for such a formulation relies on a well defined separation of time scales.

Given the assumption that discreteness effects are associated primarily with interactions between particularly proximate particles (cf. Chandrasekhar's (1941) "nearest neighbour approximation"), one would anticipate that the time scale associated with any given interaction should be exceedingly short, much shorter than a characteristic crossing time, or any other time scale of physical interest. This would suggest that these encounters can be idealised as a sequence of essentially instantaneous pointlike events. To the extent that successive events are uncorrelated, it is then natural to view them as a succession of random impulses, *i.e.*, noise. However, allowing only for such noise would lead to the physically unrealistic conclusion that discreteness effects systematically pump energy into the orbits. This difficulty can be remedied by augmenting the noise with a dynamical friction, which serves systematically to remove energy from the orbits. The strength and form of the friction and noise must be related by a fluctuation-dissipation theorem that ensures an appropriate energy balance.

This chain of reasoning, which can be justified rigourously for interactions via shorter range forces (cf. Bogoliubov 1946), is completely consistent with Chandrasekhar's original calculations of the effects of collisions in a binary encounter approximation. In particular, the diffusion coefficient computed in Chandrasekhar (1942) and the coefficient of dynamical friction computed in Chandrasekhar (1943a) are indeed related in precisely the way required by a fluctuation-dissipation theorem, a fact which figures heavily in Chandrasekhar's subsequent analyses (cf. Chandrasekhar 1943b, 1943c).

Strictly speaking, the logarithmic divergences arising in the limit of large impact parameter suggest that the assumed time scale separation may not be as clean as one might like to hope. However, to the extent that the dominant contributions to the friction and noise arise from relatively small impact parameters, the time scale separation is in fact justified, at least approximately.

Although less well known to astronomers, it is oftentimes true that weak couplings to an external environment can also be approximated by friction and noise. Consider an arbitrary Hamiltonian system, coupled to an external environment, assumed characterised by a time-independent Hamiltonian, that can be approximated as a stable bath. Here the fact that the environment is acting as a stable bath means that each individual degree of freedom is only weakly coupled to the system. This implies that the couplings only serve to trigger stable linear oscillations of the bath about some unperturbed configuration, and that they must be linear in the bath variables (Caldeira & Legett 1983a,b).

In this case, one can extract (cf. Ford *et al* 1965, Zwanzig 1973, Habib & Kandrup 1992) an exact nonlocal equation for the evolution of the system which satisfies an exact fluctuation-dissipation theorem. More precisely, by eliminating explicit reference to the degrees of freedom of the bath, one obtains an evolution equation in which the forces acting on the system at time $t$ depend on the state of the system at earlier times $s < t$. Significantly, this equation is no longer Hamiltonian.

There is no guarantee *a priori* that this nonlocal description can be approximated as a Markov process, with delta-correlated noise and a coefficient of dynamical friction that depends only on the orbital characteristics at time $t$. Indeed, a great deal of work in statistical physics has focused on the problem of determining precisely when such an approximation is justified (cf. Lebowitz & Rubin 1963, Mazur & Braun 1964, Barone & Caldeira 1991, and references contained therein). Physically, one anticipates that a Markov approximation is justified if the bath contains a large number of degrees of freedom characterized by a broad range of smoothly varying frequencies.

When considering the effects of an external environment, one should envision a large number of degrees of freedom coupled to an individual point mass star moving in a given galaxy. In analysing the effects of these many degrees of freedom, one might expect to see a broad range of relevant time scales, extending from very short, $t \ll t_{cr}$, to relatively long, $t \geq t_{cr}$. Slow effects, acting coherently on time scales $\sim t_{cr}$, should presumably be modelled as systematic time-dependent corrections to the unperturbed equations of motion. However, more rapid effects, proceeding on time scales $\ll t_{cr}$, may instead act in an irregular, near-random fashion when viewed over somewhat longer time scales. To the extent that this is true, one would anticipate that they may be modelled approximately as Markovian noise.

A correct analysis of the influences associated with an external environment must allow for both high and low frequency effects. The high frequency effects probably may be modelled, at least approximately, as friction and noise using the types of techniques described in this paper. However, low frequency effects require substantially different techniques, which will be considered elsewhere (Abernathy *et al* 1994).

This sort of approach leads generically to multiplicative noise, *i.e.*, noise which depends on the position and velocity of the orbit, and hence, because of the fluctuation-dissipation theorem, a coefficient of dynamical friction that also depends on position and velocity. However, following the earliest work of Chandrasekhar (1942b) the cal-culations summarised in this paper assume additive noise, which is independent of position and velocity. This paper focuses on the basic physical processes, considering relatively simple model potentials and friction and noise of a particularly simple form. The consideration of more general forms of friction and noise, in the context of more realistic galactic potentials, e.g., of the forms considered by Wozniak (1993), will be considered in a later paper (Habib *et al* 1994b).

Conventional wisdom would of course say that very weak friction and noise, reflecting, *e.g.*, the effects of collisionality, can only have significant effects on very long time scales. However, as will be demonstrated in this paper, under certain circumstances even very weak friction and noise can produce significant effects on relatively short time scales. Thus, for example, friction and noise corresponding to a relaxation time $t_R$ as much as $10^6$ times longer than a characteristic crossing time $t_{cr}$ can have appreciable effects within a period as short as $\sim 100t_{cr}$. These effects impact both on the form of individual orbits and, even more significantly, on the statistical properties of ensembles of orbits.

Usually it is asserted that, if $t_R$ is much longer than the age of the Universe, $t_H$, the effects of friction and noise can be completely ignored. However, this assertion assumes implicitly that the collective effects associated with the bulk potential and the relaxational effects associated with the friction and noise are completely decoupled, which, in point of fact, is not always the case. The numerical results described in this paper do not suggest that weak friction and noise will be important in galaxies that are characterised by an integrable or near-integrable galactic potential, containing only regular orbits. However, if some galactic potential is far from integrable, and admits both regular and stochastic orbits, it would appear that friction and noise may be substantially more important than is generally recognised. These effects reflect an extrinsic diffusion (cf. Lichtenberg & Lieberman 1992) which arises because of the non-Hamiltonian character of the perturbing influences.

One might worry about applying these sorts of ideas to self-gravitating systems since, because of the gravothermal catastrophe, all self-gravitating systems are unstable. However, this gravothermal instability involves gross changes in the values of the collisionless invariants along individual orbits, which can only occur on the natural time scale $t_R$ associated with the friction and noise. It follows that, on time scales $t \ll t_R$, the gravothermal catastrophe cannot be triggered.

A description of a galaxy in terms of the collisionless Boltzmann equation is a mean field theory, formulated in terms of collective coordinates, which ignores both internal fluctuations associated with discreteness effects, *i.e.*, collisionality, and external fluctuations associated with couplings to a surrounding environment. However, there is good reason to believe that both these internal and external fluctuations oftentimes can be modelled as friction and noise and, consequently, it is important to explore the effects of friction and noise on self-consistent galactic potentials.

The purpose of this paper is to investigate the effects of friction and noise on ensembles of regular and stochastic orbits in strongly nonintegrable potentials, and to speculate on the implications of these effects in galactic dynamics.

This paper summarises an investigation of orbits in two strongly nonintegrable potentials which, in an appropriate sense, appear reasonably generic within the class of potentials admitting a significant measure of stochastic orbits. Section 2 defines the potentials, describes the Langevin simulations used to incorporate the effects of friction and noise, and then explains the sense in which the potentials appear to be generic. It concludes by providing some motivation for the consideration of such nonintegrable potentials admitting global stochasticity.

Section 3 examines the effects of friction and noise on individual orbits. This examination confirms the conventional wisdom that, when viewed in energy space, friction and noise simply serve to induce a classical diffusion process which proceeds on a relaxation time $t_R$, independent of whether the unperturbed orbit is regular or stochastic. However, when viewed in configuration or velocity space, friction and noise have significantly more complicated effects which depend critically on whether the unperturbed orbit is regular or stochastic. For the case of regular orbits, friction and noise lead to a perturbed trajectory that only diverges from the unperturbed Hamiltonian trajectory as a power law in time. However, for the case of stochastic orbits the divergence is instead exponential, with a rate that is related to the Lyapunov exponent for the unperturbed orbit.

Despite these differences there is one common feature observed for both regular and stochastic orbits. Specifically, by comparing multiple noisy realisations of a single initial condition to the deterministic Hamiltonian trajectory generated from the same initial condition, one observes a simple scaling behaviour in terms of the amplitude of the friction and noise.

The fact that friction and noise can significantly alter the form of an individual stochastic orbit on very short time scales does not necessarily imply that it will also change the statistical properties of ensembles of stochastic orbits. One might expect instead that, on short times $\ll t_R$, friction and noise serve simply to continually deflect one stochastic orbit with given energy to another stochastic orbit, with almost the same energy and the same statistical properties. However, an analysis of the effects of friction and noise on ensembles of stochastic orbits, described in Section 4, shows that, in certain cases, the statistical properties can be altered dramatically. These changes reflect diffusion through cantori surrounding islands of regularity embedded in the stochastic sea and, as such, may have significant implications for galactic models that include sticky, or confined, stochastic orbits (cf. Sellwood & Wilkinson 1993).

Sections 3 and 4 focus exclusively on the calculations effected for the potentials described in Section 2. Section 5 ends by summarising the principal conclusions derived from these calculations and then speculating on the potential implications for galactic dynamics.

# 2 A Formulation of the Problem

## 2.1 Description of the Potentials and the Calculations

The calculations in this paper focus on the effects of friction and noise on orbits in two simple, two degree of freedom Hamiltonian systems. The first of these corresponds to a Hamiltonian of the form

$$H = \frac{1}{2}\left(v_x^2 + v_y^2\right) + \frac{1}{2}\left(x^2 + y^2\right) + x^2 y - \frac{1}{3}y^3 + \frac{1}{2}x^4 + x^2y^2 + \frac{1}{2}y^4$$
$$+x^4 y + \frac{2}{3}x^2y^3 - \frac{1}{3}y^5 + \frac{1}{5}x^6 + x^4y^2 + \frac{1}{3}x^2y^4 + \frac{11}{45}y^6, \qquad (1)$$

where $x$ and $y$ represent configuration space coordinates and $v_x$ and $v_y$ the corresponding velocities. The potential in eq. (1) can be derived from the well known Toda (1967) potential by expanding in a power series in $x$ and $y$ and then truncating at sixth order. Had one truncated instead at third order, he or she would be led to the Hénon-Heiles (1964) potential. As is well known (cf. Lichtenberg & Lieberman 1992) the full Toda potential leads to a completely integrable system. However, aside from the second order truncation, which corresponds to two independent oscillators, all finite order truncations lead to systems which include both regular and stochastic orbits (Yoshida *et al* 1988). At very low energies this sixth order truncation admits few, if any, stochastic orbits but, above an energy $E \approx 0.80$ both regular and stochastic orbits coexist. As the energy is increased above this value, there is a very rapid initial increase in the relative size of the stochastic phase space regions. However, for energies larger than $E \approx 24$ the rate of growth decelerates, although the relative size appears to increase monotonically (cf. Contopoulos and Polymilis 1987).

The second potential corresponds to a Hamiltonian system with

$$H = \frac{1}{2}(v_x^2 + v_y^2) - (x^2 + y^2) + \frac{1}{4}(x^2 + y^2)^2 - \frac{1}{4}x^2y^2. \qquad (2)$$

This potential, a special case of the dihedral $D4$ potential introduced by Armbruster *et al* (1989), is a squared version of the Mexican hat potential. At very low energies all orbits in this potential appear to be regular. However, above a certain critical value, one again sees a coexistence of regular and stochastic orbits. The relative size of the stochastic regions originally increases to a maximum value at $E \approx 1.0 - 1.5$ and then begins to decrease, although stochastic orbits seem to persist at arbitrarily high energies.

Both of these potentials are similar in that they admit a substantial amount of stochasticity, even though they are characterised by different shapes and symmetries. The truncated Toda potential is approximately triangular, with a $2\pi/3$ rotation symmetry, while the dihedral potential is roughly square, with a $\pi/2$ rotation symmetry. It is straightforward to verify that, for both potentials, the stochastic regions of any given energy are connected, so that every stochastic orbit of a given energy eventually probes the same phase space region. This implies, in particular, that every stochastic orbit of energy $E$ is characterised by the same Lyapunov exponent, $\chi(E)$.

These potentials were obviously not chosen with the aim of modelling any particular class of galaxy. Rather, in the pioneering of spirit of Hénon and Heiles (1964), the aim was to select potentials which are generic in their basic properties, but convenient computationally, so that large numbers of computations can be performed. The basic conclusions described in this paper are robust in that they hold for both potentials considered here, and most likely, for many other nonintegrable potentials as well. In particular, there are only three obvious characteristics that these potentials share. First, for a broad range of energies, the phase space contains substantial measures of both regular and stochastic orbits. Second, each potential admits a discrete symmetry. Third, in each case the constant energy hypersurfaces are compact, *i.e.*, of finite volume.

As discussed in more detail below, the first of these features is crucial in what follows. However, the second is probably not all that important. Indeed, there is a good deal of numerical evidence (cf. Udry & Pfenniger 1988) that would suggest that, although breaking these, or any other, discrete symmetries may tend to increase the relative measure of the stochastic phase space regions, the qualitative features of the stochastic orbits that *do* exist are independent of the symmetries. The final fact that the accessible phase space is finite in volume is also crucial: only for a compact phase space can one define the notion of an invariant measure.

Mathematically, an invariant measure corresponds to a probability distribution which, if evolved into the future using the equations of motion, remains invariant (cf. Lichtenberg & Lieberman 1992). In other words, it yields a time-independent statistical equilibrium. If one is interested in constructing a self-consistent galactic model which contains stochastic orbits, using Schwarzschild's (1979) method or any variant thereof, it is important, in selecting the orbits entering into the model, to isolate on an appropriate set of time-independent building blocks. For the regular phase space regions, the fundamental building blocks correspond to collections of regular orbits trapped around the stable periodic orbits. For the stochastic phase space regions the obvious building block is the time-independent invariant measure.

In canonical coordinates the invariant measure corresponds to a uniform sampling of the accessible stochastic region. If, for example, all the orbits at a given energy are stochastic, and there are no additional conserved quantities, the invariant measure at that energy corresponds to a microcanonical distribution.

In general, one might anticipate that a generic ensemble of initial conditions corresponding to stochastic orbits will eventually evolve towards this invariant measure. In this case the Ergodic Theorem implies an equivalence of temporal and phase space averages, provided that the phase space average is computed with respect to the invariant measure. It has been shown for various potentials (cf. Kandrup & Mahon 1994a, Mahon *et al* 1994) that, in a coarse-grained sense, generic ensembles evidence an exponential approach in time towards a distribution which, to within statistical uncertainties, appears to be approximately constant in a time-averaged sense.

It is not, however, obvious that this approximately time-invariant distribution corresponds to the true invariant measure, and indeed, there is substantial numerical evidence that it does not (cf. Lieberman & Lichtenberg 1972, Mahon *et al* 1994). Differences between this approximate invariant measure and the true invariant measure are particularly pronounced for energies where the stochastic sea contains large islands of regularity surrounded by cantori. These cantori, which are fractured KAM tori (Mather 1982), serve to repel stochastic orbits on short time scales (cf. MacKay *et al* 1984a,b; Lau *et al* 1991), although these orbits can pass through the cantori and become trapped around the regular regions on sufficiently long time scales.

Translated into the language of galactic dynamics, the approach towards a near-invariant measure corresponds to a systematic evolution towards a statistical quasi-equilibrium. However, the presence of the cantori implies that this quasi-equilibrium need not correspond to a true equilibrium, even in the context of a deterministic evolution in a smooth bulk potential. If this quasi-equilibrium is evolved for longer time scales, a slow, albeit systematic, evolution may be observed.

The combined effects of friction and noise were modelled here by performing Langevin simulations. These involve solving evolution equations of the form

$$\frac{d\mathbf{r}}{dt} = \mathbf{v} \qquad \text{and} \qquad \frac{d\mathbf{v}}{dt} = -\nabla\Phi - \eta\mathbf{v} + \mathbf{F}. \tag{3}$$

Here $\Phi(x, y)$ denotes the smooth potential, $\eta\mathbf{v}$ is the dynamical friction, and $\mathbf{F}$ represents the noise associated with the random kicks. If one neglects $\eta\mathbf{v}$ and $\mathbf{F}$, one recovers the deterministic Hamiltonian systems associated with eqs. (1) and (2).

The noise was modelled as delta-correlated additive white noise with zero mean, related to the constant friction $\eta\mathbf{v}$ via a fluctuation-dissipation theorem (Chandrasekhar 1943c). Because it is a random quantity, $\mathbf{F}$ is characterised completely by its moments. The assumption of zero mean implies that the first moment $\langle \mathbf{F}(t) \rangle = 0$ for all times $t$. The value of the second moment is fixed by the fluctuation-dissipation theorem which relates the random force to the dynamical friction in terms of a characteristic "temperature" or mean squared velocity $\Theta$, *i.e.*, component by component

$$\langle F_i(t_1) F_j(t_2) \rangle = 2\Theta\eta\delta_{ij}\delta_D(t_1 - t_2). \tag{4}$$

Here $\delta_{ij}$ denotes a Kronecker delta and $\delta_D(t_1 - t_2)$ denotes a Dirac delta.

Numerical solutions were generated by solving a discretised version of eq. (3) using a numerical algorithm that generates the proper first and second moments (Griner *et al* 1988). Most of the simulations were effected with time step size $h = 10^{-3}$. It was confirmed that a smaller time step $h = 10^{-4}$ leads to statistically indistinguishable results.

For each initial condition, an unperturbed Hamiltonian trajectory was first computed. Then multiple noisy realisations were effected, and the results of these simulations combined to extract statistical properties. This involved two different types of analysis. First, the noisy orbits were compared with the deterministic Hamiltonian trajectories to extract the first and second moments for the deviations in configuration, velocity, and energy space. Here, for example,

$$\langle \delta x \rangle = \frac{1}{N} \sum_{i=1}^{N} (x_{det} - x_i) \tag{5}$$

and

$$\langle |\delta x|^2 \rangle \equiv \delta x_{rms}^2 = \frac{1}{N} \sum_{i=1}^{N} (x_{det} - x_i)^2, \tag{6}$$

where the sum extends over the $N$ different noisy realisations. Second, the noisy data were binned to obtain a coarse-grained representation of the phase space density associated with the orbits.

Two different types of simulations were performed. First, to examine the effects of friction and noise on individual orbits, multiple noisy realisations of individual initial conditions were effected, and the results of the different initial conditions examined separately. Most of these experiments involved 50 different realisations, which were analysed to extract the aforementioned moments. However, a few of the experiments were performed with much larger numbers of realisations, up to 4000, such larger numbers facilitating the possibility of constructing coarse-grained representations of the phase space density.

The second type of simulation, effected only for the truncated Toda potential, involved 100 noisy realisations each of ensembles of 400 stochastic initial conditions so chosen as to sample the deterministic near-invariant measure described above (cf. Kandrup & Mahon 1994a,b). This latter set of experiments was performed for $E = 30$ and 75 with $\Theta = E$ and $\eta = 10^{-9}$, $10^{-6}$, and $10^{-4}$. A single experiment was done with $E = \Theta = 10$ and $\eta = 10^{-6}$. Two sets were done with $E = 75$ and $\eta = 10^{-6}$ to verify that different seeds lead to statistically identical conclusions. Each of these latter experiments took approximately 250 node hours on the Los Alamos CM-5.

The motivation for the simulations described in this paper is to provide insights into the structure and evolution of galaxies, which are typically only $\sim 100$ characteristic crossing times $t_{cr}$ in age. Since, for the Hamiltonian systems (2) and (3), the typical crossing time is itself of order unity in absolute units, all the integrations were therefore restricted to a total time $t \leq 140$. If the friction and noise are intended to model the effects of collisionality, the natural choice for the strength of the friction is easily estimated. For a galaxy like the Milky Way, the collisional relaxation time $t_R \sim 10^{14}$ yr is some six orders of magnitude larger than a characteristic crossing time, $t_{cr} \sim 10^8$ yr, so that it is natural to select a value $\eta \sim t_{cr}/t_R \sim 10^{-6}$. If the friction and noise are instead intended to model external influences, the appropriate value of $\eta$ is less well determined: it could be larger. The simulations reported here explored a broad range of values, $10^{-12} \leq \eta \leq 10^{-3}$. The natural magnitude of $\Theta$ is fixed by its interpretation as a typical mean squared velocity, which should presumably be comparable to the typical values observed for the deterministic orbits. This was achieved by allowing for a more restricted range of values, $0.1 \leq E/\Theta \leq 10.0$.

It is well known that, on the natural time scale $t_R \sim \eta^{-1}$, a time typically much longer than the age of the Universe, friction and noise try to drive a systematic evolution towards a thermal distribution. This is not the point of interest here. Rather, the question is whether friction and noise can also induce significant effects on much shorter time scales, $\leq t_H$.

These sorts of Langevin simulations are closely related to numerical solutions to a Fokker-Planck equation for the evolution of orbits in a fixed background potential (cf. Chandrasekhar 1943c, van Kampen 1981). Indeed, performing Langevin simulations for a given initial condition $\{\mathbf{r}(0), \mathbf{v}(0)\}$ addresses the same physical situation as solving a Fokker-Planck equation for a probability distribution $f(t)$ with initial condition $f(t{=}0) = \delta_D(\mathbf{r} - \mathbf{r}(0))\delta_D(\mathbf{v} - \mathbf{v}(0))$. Fokker-Planck simulations have the advantage that they yield directly the full distribution function, from which all the moments can be extracted. However, Langevin simulations have the advantage that they allow one to examine individual noisy orbits. From a numerical standpoint, Langevin equations are easier to deal with as one has to solve a set of stochastic ordinary differential equations rather than a partial differential equation. The advent of modern supercomputers possessing large amounts of memory has made possible the solution of Fokker-Planck equations in two spatial dimensions with good resolution (cf. Habib and Ryne 1994). However, extending these methods to three dimensions still remains out of reach at present.

## 2.2 Nonintegrable Galactic Potentials

At this stage, it is perhaps worthwhile to review the motivation in galactic dynamics for the consideration of potentials that admit stochastic orbits.

A knowledge of the mass distribution of some galaxy, and hence the gravitational potential, does not in general determine uniquely the form of the phase space distribution (cf. Hunter 1994) and, as such, one might think that, even if the potential admits stochastic orbits, one can construct equilibria that involve only regular orbits. This is not so. Specifically, one can show that if some time-independent mass density corresponds to a potential that admits both regular and stochastic orbits, any self-consistent phase space distribution which yields that density must also contain both regular and stochastic orbits.

That this is so is easily seen. Suppose that the equilibrium distribution $f_0$ is a function of $n$ ($\leq 3$) isolating integrals. By restricting attention to fixed values of the isolating integrals, one is restricted to a $(6-n)$-dimensional phase space hypersurface. A necessary condition for equilibrium is that $f_0$ sample this hypersurface uniformly. In general, this hypersurface will contain both regular and stochastic regions, and the regular regions may be divided into different orbit types. Equilibrium requires that all these different types be sampled in such a fashion as to yield a constant population of the hypersurface: If two different types of regular orbits exist, they must both be sampled; if regular and stochastic orbits coexist, they too must both be sampled.

It follows that one cannot select a potential that incorporates stochastic orbits but simply choose to avoid populating the stochastic regions. One must either pick a potential which admits only regular orbits or consider a more general potential, including both regular and stochastic orbits.

Several workers (cf. Contopoulos 1993) have discovered that allowing for stochastic orbits seems to be useful in the construction of self-consistent models by helping to support various structures, *e.g.*, connecting bars to spirals in a barred spiral galaxy (Kaufmann 1993). These workers would agree with the conventional wisdom (cf. Bin-ney 1978a,b, de Zeeuw 1985) that regular orbits provide a "skeleton" for the mass distribution that generates the potential. However, they would invoke stochastic orbits to populate various phase space regions where regular orbits do not exist, *e.g.*, near resonances such as corotation. This intuition is corroborated by detailed analysis of $N$-body simulations (cf. Sparke & Sellwood 1987, Pfenniger & Friedli 1991), that suggest that stochastic orbits do indeed play a role in supporting a bar.

It should also be observed that calculations of orbits in smooth analytic potentials suggest that even relatively small symmetry-breaking corrections, superimposed on some idealised unperturbed model, can dramatically increase the overall abundance of stochastic orbits, as measured, *e.g.*, by the Kolmogorov entropy. For example, Udry & Pfenniger (1988) and Hassan and Norman (1990) have shown that a central mass concentration, *e.g.*, reflecting the effects of a supermassive black hole, inserted into a galactic model can destabilise the central box orbits and induce a large amount of stochasticity. Similarly, Udry & Pfenniger (1988) have shown that perturbing a plane symmetric model by allowing for $l = 3$ or 4 harmonic $P_l(\cos\theta)$ perturbations, with amplitudes comparable to those observed for galaxies in rich cluster environments (cf. Zepf & Whitmore 1993), can greatly increase the relative measure of stochastic orbits. Even changing the form of the symmetry can have significant effects. For example, Athanassoula (1990) has shown that, by making a bar more squarelike, and less elliptical, in shape one seems to increase the relative abundance of chaotic orbits.

At present, one does not know whether real systems are characterised by a mass density that yields an integrable, or near-integrable, mean field potential admitting only regular orbits, or whether instead the mean field potential is strongly nonintegrable and admits stochastic orbits. It may be true that galactic potentials are all nearly integrable, but that astronomers have not yet been clever enough to identify the "right" potentials. However, there is no proof.

It would seem extremely difficult, if not impossible, to determine directly from observations of a galaxy whether or not the mean field potential admits stochastic orbits. Even in the best of circumstances, mapping out the overall mass distribution is not easy, and one knows (cf. Lichtenberg and Lieberman 1992) that, despite the KAM theorem, there is a sense in which two potentials relatively similar in appearance can admit very different orbit types. The integrable Toda potential and its nonintegrable truncations provide a particularly simple illustration thereof. Another example is provided by the logarithmic potential, and various perturbations thereof, as considered by Binney and Tremaine (1987).

The numerical evidence from $N$-body simulations is also inconclusive. It is true that many orbits in a simulation can often be identified unambiguously as correspond-ing to specific regular types (cf. Barnes & Hernquist 1991), but for other orbits this is not possible (cf. van Albada 1986). At present, it is not completely clear whether this reflects (a) discreteness effects, which may be more important in small $N$-body simulations than in large galaxies, (b) time-dependent variations in the mean field potential, reflecting the fact that the system has not reached a strict time-independent equilibrium, or (c) the actual existence of real stochastic orbits.

Theoretically, the situation is again unclear. From the viewpoint of nonlinear dynamics *per se*, there is no reason *a priori* to assume either that galactic potentials are all integrable or near-integrable, admitting only regular orbits, or that they are far from integrable, admitting both regular and stochastic orbits. Integrable potentials are of course a set of measure zero in the set of all potentials, but near-integrable potentials which, by virtue of the KAM Theorem, admit qualitatively similar orbits, need not be of zero measure.

If it is true that self-gravitating systems like galaxies tend generically to evolve towards a mass distribution corresponding to a mean field potential that admits only regular orbits, this must be because of some dynamical principle which has not yet been identified unambiguously.

Implicit in the construction of self-consistent equilibria is the assumption that the evolution of the system is determined by the collisionless Boltzmann equation, which is believed to describe correctly the short time behaviour of the $N$-body problem in the limit $N \to \infty$. However, one knows (a) that this collisionless Boltzmann equation is an infinite-dimensional Hamiltonian system (cf. Morrison 1980) and (b) that an evolution described by this equation is strongly constrained by the infinite number of conserved quantities associated with Liouville's Theorem. If these constraints were sufficient to imply that the collisionless Boltzmann equation is an integrable system, this might be interpreted as evidence that, in the limit of large $N$, the $N$-body problem is nearly integrable. However, the fact that all these constraints are ultra-local, *i.e.*, involving no derivatives of the distribution function, would suggest strongly (cf. Morrison 1987) that, unlike the Korteweg-de Vries and other similar equations, the collisionless Boltzmann equation is *not* integrable.

To summarise, there is at present no compelling evidence to exclude the possibility of stochastic orbits in galactic mean field potentials, and there *are* indications that such orbits may in fact exist. It is therefore important to explore the implications of the hypothesis that galactic potentials do admit global stochasticity and, in particular, to investigate the effects of friction and noise in such systems.

# 3 The Effects of Friction and Noise on Individual Orbits

Fundamentally, the deterministic Hamiltonian systems (1) and (2) admit only two different classes of orbits, namely regular orbits, which are characterised by a vanishing Lyapunov exponent $\chi$, and stochastic orbits, for which $\chi$ is nonvanishing. However, when viewed on relatively short time scales, the stochastic orbits divide in turn into two more or less distinct classes, namely filling stochastic orbits that travel unimpeded throughout most of the stochastic sea, and confined, or sticky, stochastic orbits, that appear trapped in the neighbourhood of an island of regularity.

All the regular orbits are of course separated from all the stochastic orbits by invariant KAM tori, so that the distinction between regular and stochastic is absolute and robust: without introducing sizeable non-Hamiltonian perturbations, one orbit class cannot be transformed to another. However, the two different classes of stochastic orbits are only separated by cantori, *i.e.*, fractured KAM tori that contain a Cantor set of gaps. This means that transitions between filling and confined stochasticity can occur via so-called intrinsic diffusion, although the characteristic time scale associated with this process is typically $t \gg 100$.

On time scales short compared with the intrinsic diffusion time scale, a deterministic evolution leads only to a minimal mixing of filling and confined stochastic orbits, so that it is meaningful as a practical matter to speak of three distinct orbit classes, namely (1) regular, (2) confined stochastic, and (3) filling stochastic. This implies in turn that, when studying the effects of friction and noise, it is natural to consider separately the effects on these three different classes of orbits, although one must also contemplate the possibility of transitions between these different classes.

## 3.1 Friction and Noise in Energy Space

Viewed in energy space, one finds that, for both the truncated Toda and $D4$ potentials, the effects of friction and noise serve simply to induce a classical diffusion process, in which the root mean squared $\delta E_{rms}$ grows as $t^{1/2}$. One might anticipate physically that the effects of the friction should only enter through the dimensionless combination $\eta t$, and that the second moment should be linear in the combination $\Theta\eta$. It would then follow on dimensional grounds that $\delta E_{rms}^2 \propto E\Theta\eta t$. In point of fact, the data are well fit by a simple scaling relation

$$\delta E_{rms}^2 = A^2(E) E\eta\Theta t, \tag{7}$$

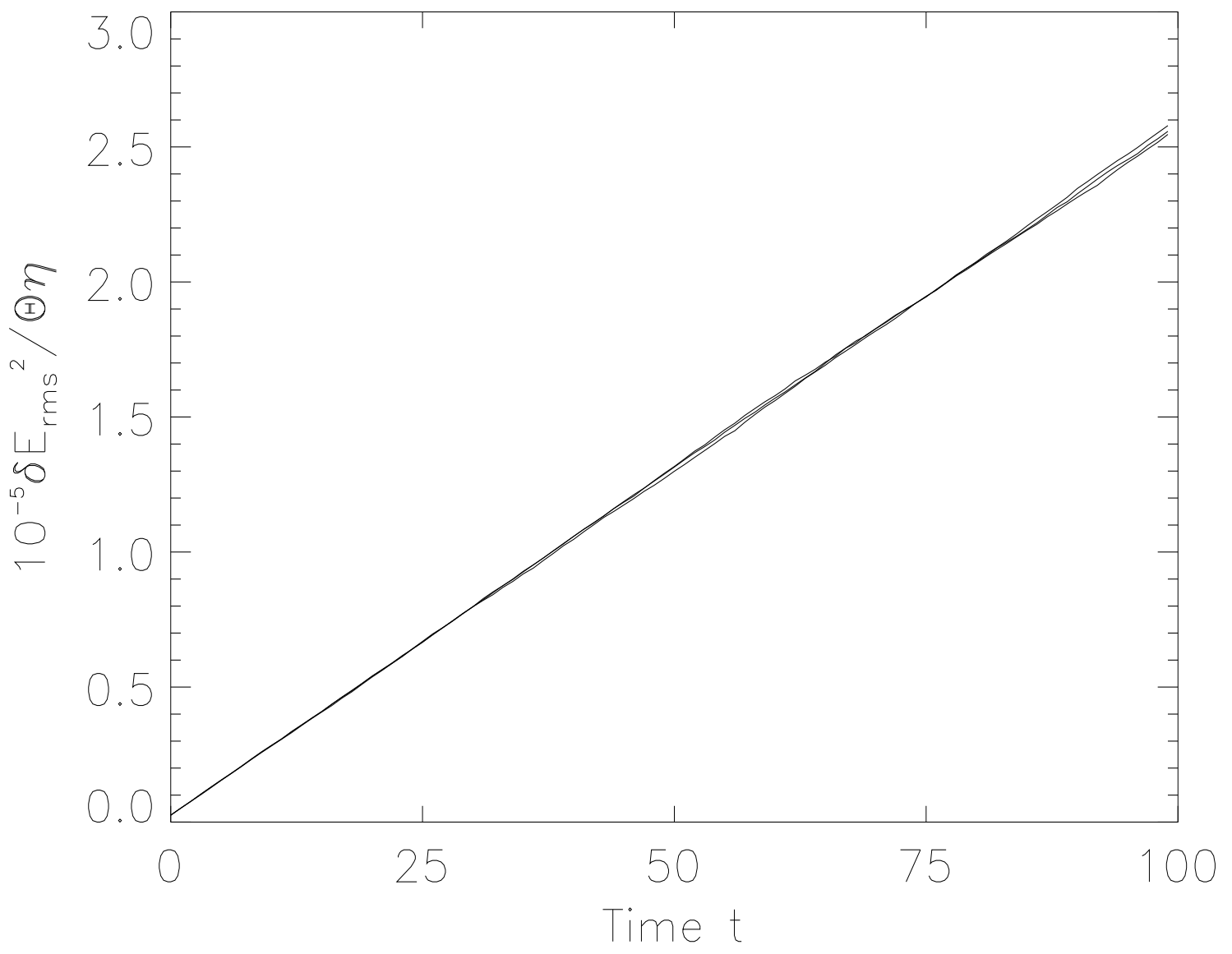

Figure 1: The ratio $\delta E_{rms}^2/\Theta\eta$, generated from a single 400 orbit ensemble of stochastic initial conditions in the truncated Toda potential, with $E = \Theta = 30$ and variable $\eta = 10^{-9}$, $10^{-6}$, and $10^{-4}$.

where $A^2(E)$ is seemingly independent of $\eta$ and $\Theta$, and admits only a weak dependence on $E$. This scaling holds for all three classes of orbits, regular, confined stochastic, and filling stochastic. If the effects of friction and noise be viewed purely in energy space, one cannot distinguish between the three different classes of orbits.

Consider, for example, the truncated Toda potential. Here simulations of individual initial conditions in the range $10 \leq E \leq 100$ with variable $\Theta$ and $\eta$ exhibit a fair degree of variability ($\sim 10 - 15\%$), with best fit values of $A$ typically in the range $1.5 - 1.8$. There is no apparent residual dependence on $\Theta$ and $\eta$, although there *is* a hint that $A$ decreases slowly with increasing energy.

For the experiments averaging over an ensemble of stochastic initial conditions, the best fit values are considerably better determined. For $E = 75$, $1.68 < A < 1.71$, for $E = 30$, $1.66 < A < 1.69$, and, for $E = 10$, $1.66 < A < 1.67$. Again, there is no systematic residual dependence on $\eta$. The goodness of fit is evident from Fig. 1, which illustrates this scaling by plotting the ratio $\delta E_{rms}^2/\Theta\eta$ for the same collection of 400 stochastic initial conditions, allowing for three different values of the friction $\eta$.

However, the data do suggest that, as $E$ is decreased, the best fit value of $A$ also decreases, albeit very slowly. In the limit $E \to 0$, the orbits move in what is essentially a harmonic oscillator potential, and one would therefore anticipate an asymptotic approach to the exact oscillator result, $\delta E_{rms}^2 = 2E\eta\Theta t$, *i.e.*, $A = 2^{1/2}$. (This result, a derivation of which is presented in an Appendix, holds in the limit $\eta \ll 1$, assuming an "orbit average," so that, *e.g.*, $\sin^2 t \to 1/2$.) The data are consistent with this expectation.

For the $D4$ potential, $A$ again seems independent of $\eta$ and $\Theta$, but the dependence on $E$ is somewhat stronger. Thus, for example, the values $E = 1.0$, 6.0, 15.0, 25.0, and 50.0 yield respectively best fit values $A = 2.60\pm0.04$, $1.81\pm0.05$, $1.79\pm0.03$, $1.64\pm0.03$, and $1.72\pm0.04$. Here the largest values correlate with the energies where, as characterised by the magnitude of the Lyapunov exponent, the phase space is most chaotic (cf. Mahon *et al* 1994).

These results confirm the physical expectation (cf. Binney and Tremaine 1987) that, as far as the energy, and any other collisionless invariants, are concerned, friction and noise only induce a slow diffusion on a characteristic time scale $t_R = \eta^{-1}$, which proceeds independently of the orbit class.

## 3.2 Friction and Noise in Configuration and Velocity Space

The behaviour of the configuration and velocity space moments, $\delta x_{rms}$, $\delta y_{rms}$, $\delta v_{x,rms}$, and $\delta v_{y,rms}$, is more complicated and depends on which of the three orbit classes one considers. For the case of regular orbits, one observes that, in both the truncated Toda potential and the $D4$ potential, all four of these second moments grow as a power law in time. Alternatively, for the case of filling stochastic orbits the overall growth is approximately exponential in time. Confined stochastic orbits exhibit an intermediate behaviour, the growth starting slowly but eventually becoming exponential. This is illustrated in Fig. 2, which plots $\ln \delta r_{rms} = \ln{(\delta x_{rms}^2 + \delta y_{rms}^2)}^{1/2}$ for ten different initial conditions evolving in the $D4$ potential, with $E = \Theta = 1$ and $\eta = 10^{-6}$. The lower four curves correspond to regular orbits, and the remaining six to stochastic orbits.

Despite these differences, there *is* one universal characteristic, namely a simple scaling in terms of $\Theta$ and $\eta$. Specifically, one finds that, for both potentials, all three classes of orbits satisfy

$$\delta x_{rms}, \delta y_{rms}, \delta v_{x,rms}, \delta v_{y,rms} \propto \Theta^a \eta^b F(E,t), \tag{8}$$

where, in each case, the exponents assume the same values, $a = b = 0.50\pm0.01$. In other words, these moments exhibit the same dependence on $\eta$ and $\Theta$ as does $\delta E_{rms}$. This scaling is observed both for individual initial conditions and for ensembles of initial conditions. Figs. $3a$ and $3b$ show the second moment $\delta r_{rms}$ for one regular initial condition for the truncated Toda potential with $E = 20$, allowing for four different values of $\eta$ and three different values of $\Theta$. Figs. 4*a* and 4*b* exhibit analogous results for a confined stochastic orbit with the same energy $E = 20$. Orbits of different energy or in the $D4$ potential yield comparable data.

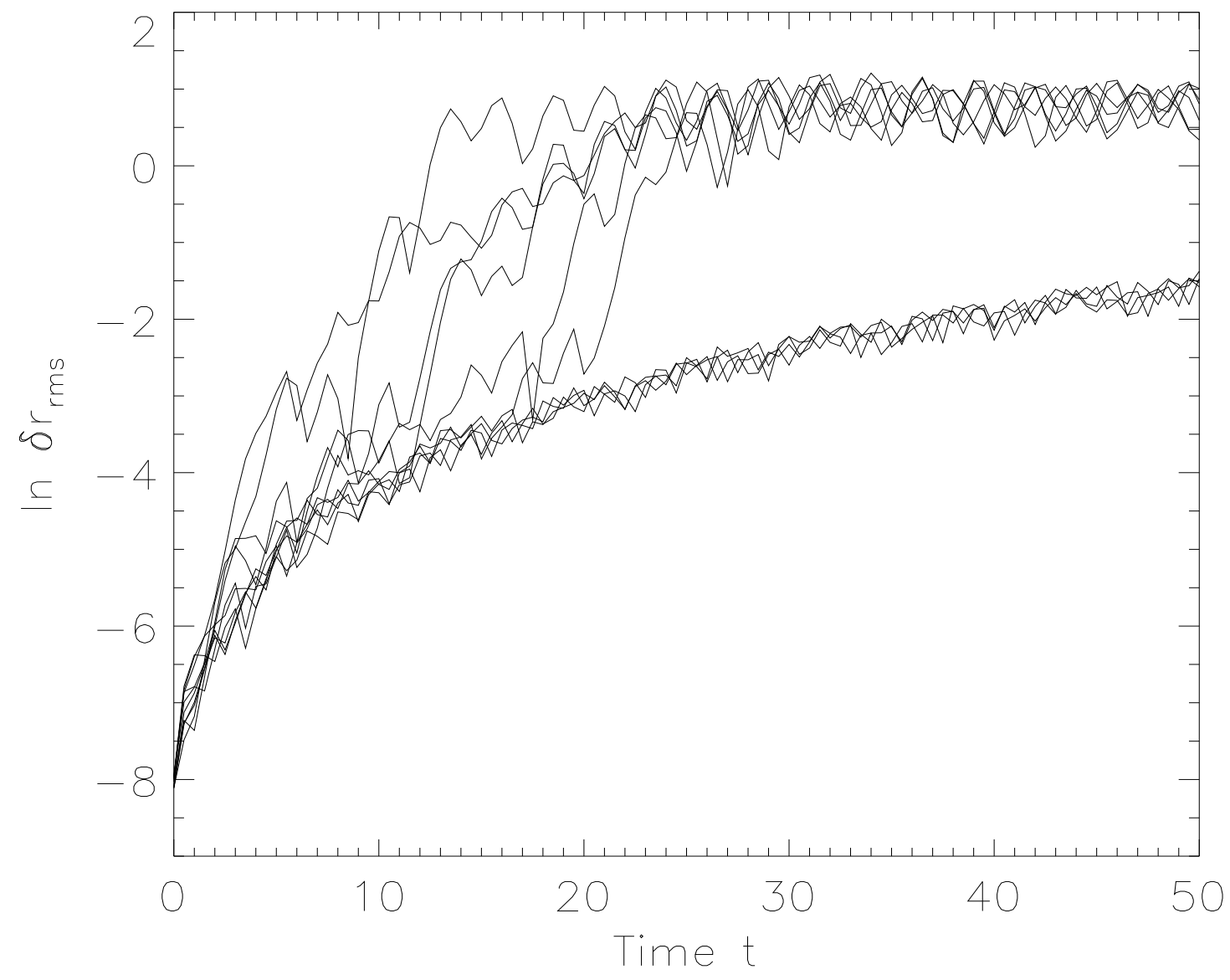

Figure 2: The quantity ln $\delta r_{rms}$, generated from ten different initial conditions in the $D4$ potential, with $E = \Theta = 1$ and $\eta = 10^{-6}$.

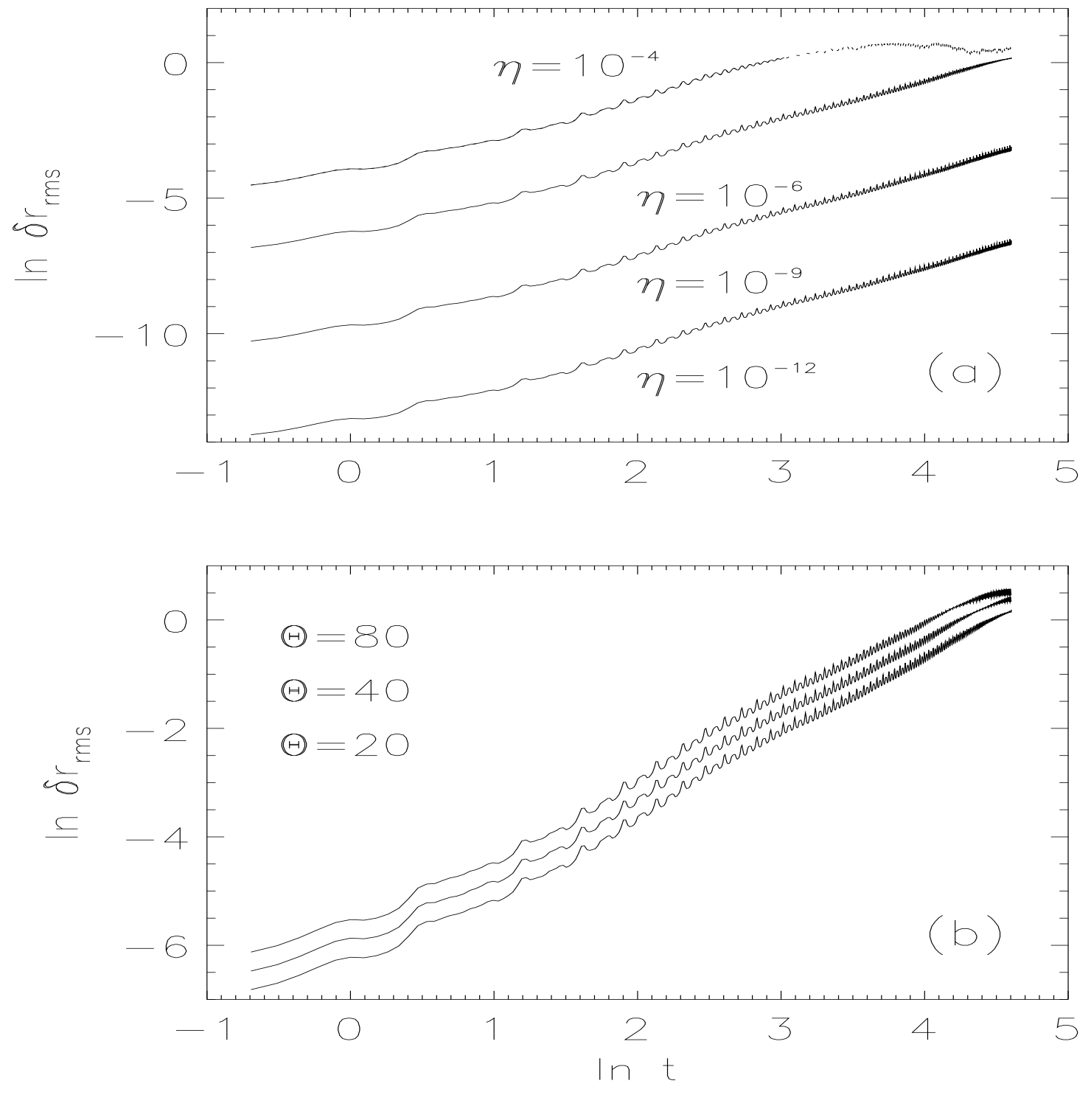

Figure 3: (a) The quantity ln $\delta r_{rms}$, generated for a single regular initial condition in the truncated Toda potential, with $E = \Theta = 20$ and variable $\eta$. (b) The quantity ln $\delta r_{rms}$, generated for a single regular initial condition in the truncated Toda potential, with $E = 20$, $\eta = 10^{-6}$, and variable $\Theta$.

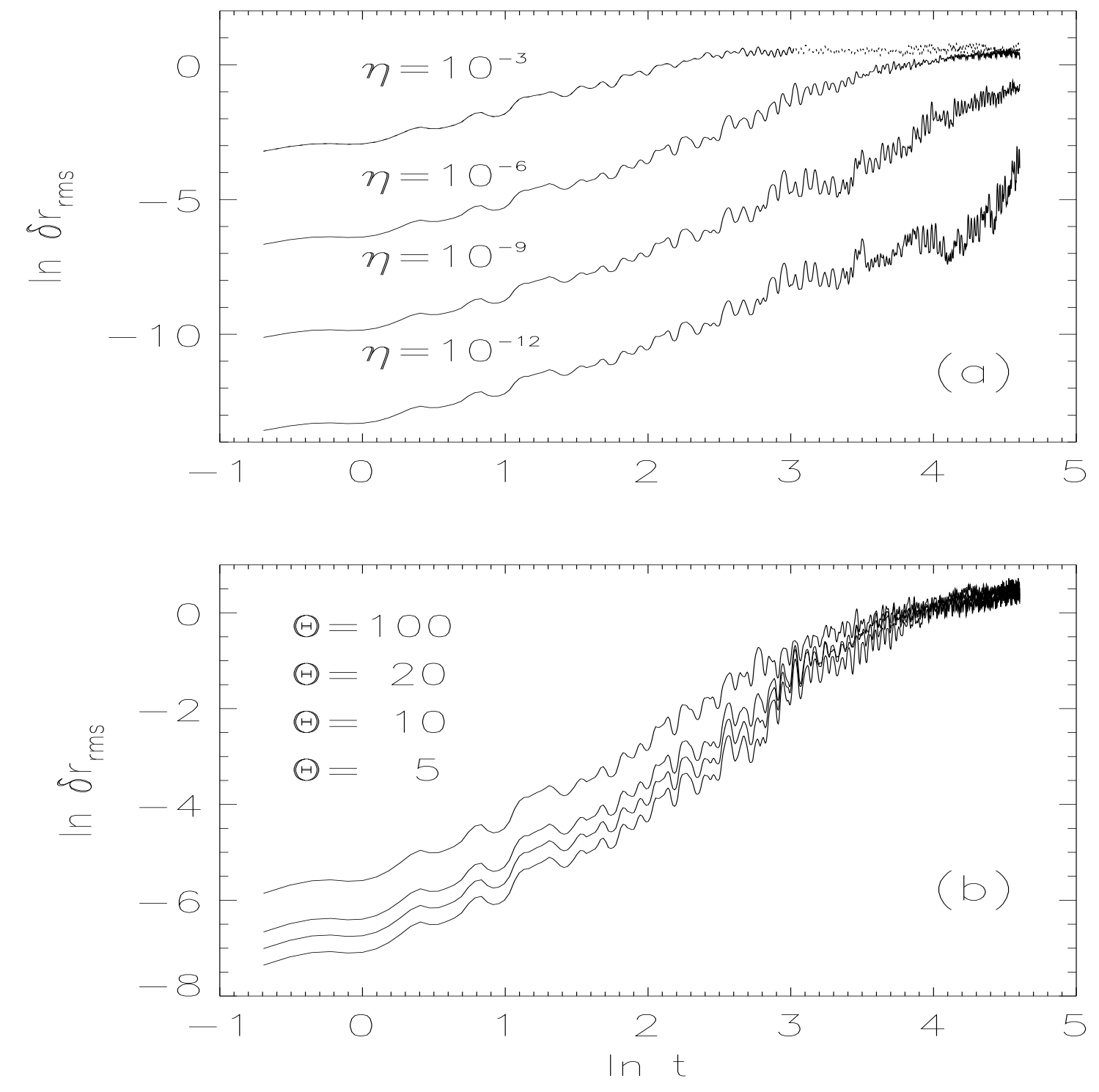

Figure 4: (a) The quantity ln $\delta r_{rms}$, generated for a single sticky initial condition in the truncated Toda potential, with $E = \Theta = 20$ and variable $\eta$. (b) The quantity ln $\delta r_{rms}$, generated for a single sticky initial condition in the truncated Toda potential, with $E = 20$, $\eta = 10^{-6}$, and variable $\Theta$.

Even for these separate initial conditions, the evolution of the moments is relatively smooth and the scaling quite distinct. However, for the ensembles of 400 stochastic initial conditions, the evolution is smoother and the scaling yet more apparent. This is illustrated in Figs. 5*a* and 5*b*, which exhibit the average $\delta r_{rms}$ as a function of time for three different values of $\eta$ for the truncated Toda potential with $E = 30$. The three solid curves represent the values $\eta = 10^{-9}, 10^{-6}$, and $10^{-4}$. The dashed and dot-dashed curves in Fig 5*b* were generated by translating the $10^{-9}$ and $10^{-6}$ curves upwards by the distances 2.3 and 5.75 predicted respectively by the scaling relation (8).

Another common feature for all three orbit types is that early on the time dependence is reasonably well fit by a power law, *i.e.*, for sufficiently early times,

$$\delta x_{rms}, \delta y_{rms}, \delta v_{x,rms}, \delta v_{y,rms} \propto f(E)\Theta^a \eta^b t^c. \qquad (9)$$

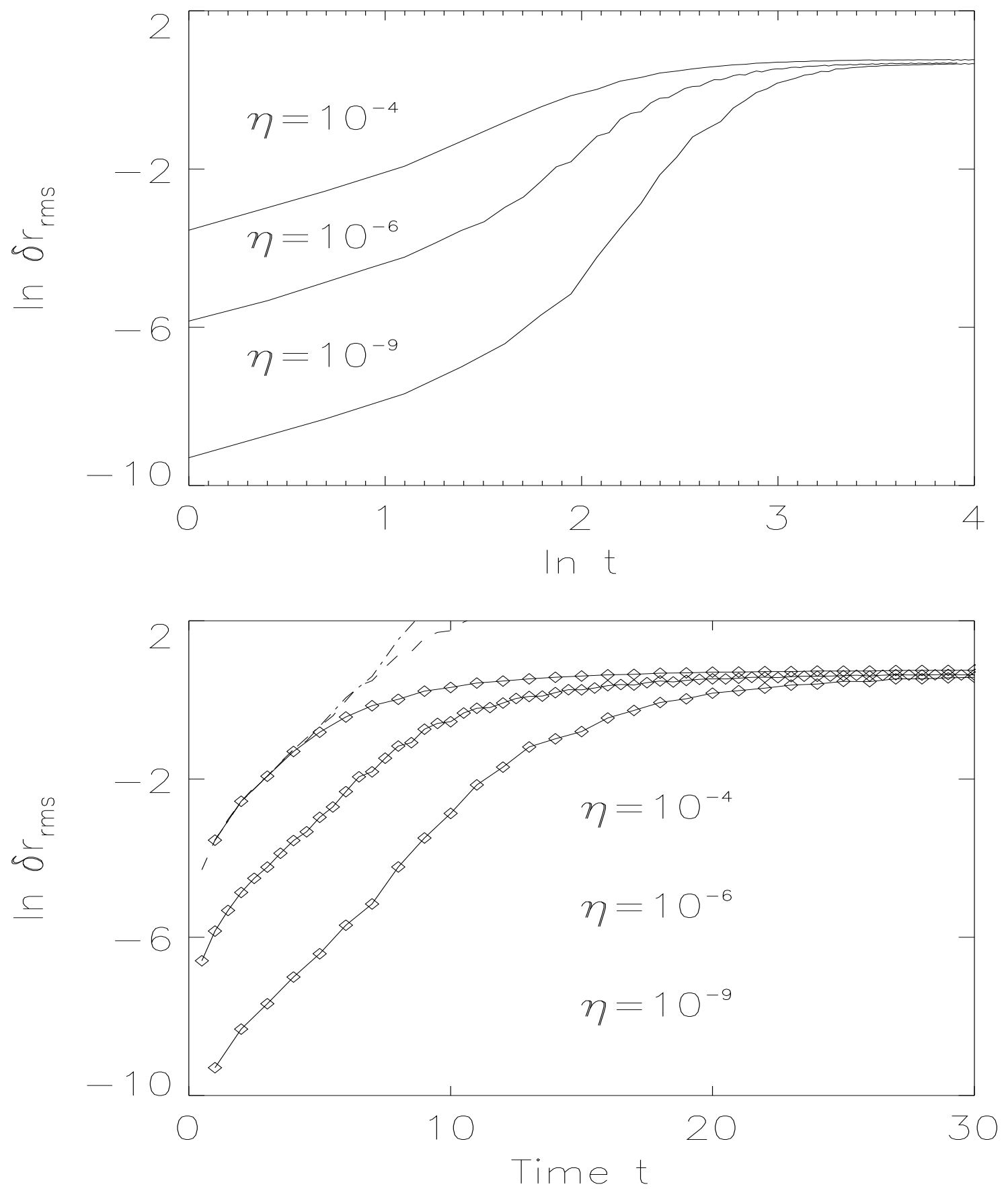

Figure 5: (a) The total perturbation $\ln\,\Delta r_{rms}$, generated from an ensemble of 400 stochastic initial conditions in the truncated Toda potential, with $E=\Theta=30$ and variable $\eta$. (b) The same data plotted as a function of $t$, rather than $\ln\,t$. The upper dashed and dot-dashed curves were generated by constant translations of ln 10 and 2.5 ln 10.

The interval over which such a fit is appropriate depends on the orbit type: filling stochastic orbits rapidly begin to show an exponential divergence, confined stochastic orbits only show this exponential divergence somewhat later, and regular orbits continue to manifest an approximate power law growth.

The best fit value of $c$ depends on the interval over which one samples: for all types of orbits, one finds that, if one fits to longer time intervals (providing of course that the perturbation has not yet become macroscopic), the best fit value of $c$ increases, even for the case of regular orbits. For both potentials, fitting for relatively short times yields a best fit value $c \approx 1.10-1.25$. Fitting to somewhat longer times yields a best fit value $c \approx 1.20-1.40$.

The observed time-dependence of these moments differs substantially from the predicted behaviour for orbits in a harmonic oscillator potential, where *e.g.*, $\delta r_{rms}^2 \propto$ $\Theta\eta t$ (cf. Chandrasekhar 1943c). To verify that this is not a numerical artifact, *e.g.*, a consequence of the finite time step $h = 10^{-3}$, an identical 400 orbit simulation was also performed for a harmonic oscillator potential, allowing for $E = \Theta = 75$ and $\eta = 10^{-6}$. The results from this simulation agree completely with theory.

These noisy simulations indicate that, at least for stochastic orbits, even very weak irregularities, modeled as friction and noise, will cause a rapid pointwise divergence away from the purely deterministic trajectory, although the noisy orbit may still have the same statistical properties. In particular, if the amplitude of the friction and noise are larger than roundoff, truncation, and other numerical errors, they will become important sooner than these other effects about which astronomers are extremely concerned (cf. Quinlan and Tremaine 1992).

## 3.3 Connections with Lyapunov Exponents

Lyapunov exponents can be defined in a number of different ways (cf. Chirikov 1979), all of which are equivalent mathematically. One particularly physical definition manifests the fact that they characterise the average rate of instability along a stochastic orbit. Specifically, for an arbitrary system one can define (cf. Bennetin *et al* 1976)

$$\chi \equiv \lim_{t\to\infty} \lim_{\delta z(0)\to 0} \frac{1}{t} \log\left(\frac{\delta z(t)}{\delta z(0)}\right), \tag{10}$$

in terms of an asymptotic, $t \to \infty$ limit, where $\delta z(0)$ and $\delta z(t)$ denote respectively the Euclidean phase space deviations of two nearby orbits at times 0 and $t$.

Given this interpretation of $\chi$, it is also natural to define local, or short time, Lyapunov characteristic numbers $\chi(\Delta t)$ which characterise the average instability over finite time intervals $\Delta t$ (cf. Grassberger *et al* 1988, Sepúlveda *et al* 1989). Specifically,

$$\chi(\Delta t) \equiv \lim_{\delta z(0)\to 0} \frac{1}{\Delta t} \log\left(\frac{\delta z(\Delta t)}{\delta z(0)}\right). \tag{11}$$

Such short time Lyapunov exponents can be used to characterise the degree of instability both for different segments of a single stochastic orbit and, of more direct relevance for galactic dynamics (cf. Kandrup & Mahon 1994b), for various ensembles of orbits, *e.g.*, ensembles sampling the invariant measure associated with deterministic Hamiltonians like (1) and (2).

If one interprets the exponentially growing deviation between noisy and deterministic trajectories as resulting from an instability triggered by the friction and noise (cf. Pfenniger 1986, Kandrup & Willmes 1994), one would anticipate that, on the average, the second moments $\delta x_{rms}$, $\delta y_{rms}$, $\delta v_{x,rms}$, and $\delta v_{y,rms}$ should grow at rates that are closely related to the Lyapunov exponent. Suppose, for example, that one samples the invariant measure to extract an ensemble of stochastic initial conditions, and tracks the total configuration space perturbations

$$\Delta x^2_{rms} = \sum_{i=1}^{N} \delta x^2_{i,rms} \qquad \text{and} \qquad \Delta y^2_{rms} = \sum_{i=1}^{N} \delta y^2_{i,rms}, \tag{12}$$

or the corresponding velocity space perturbations $\Delta v^2_{x,rms}$ and $\Delta v^2_{y,rms}$, which sum over the different initial conditions. One might then anticipate an exponential growth, $\Delta x_{rms}$, $\Delta y_{rms}$, $\Delta v_{x,rms}$, $\Delta v_{y,rms} \propto \exp(+\Lambda t)$, where $\Lambda(E)$ is comparable in magnitude to $\chi(E)$.

This hypothesis was tested in detail for the simulations involving ensembles of 400 initial conditions, and the basic picture confirmed. In particular, it was discovered that the growth of each of the second moments $\Delta x_{rms}$, $\Delta y_{rms}$, $\Delta v_{x,rms}$, and $\Delta v_{y,rms}$, is approximately exponential, and that the rates are comparable. The relevant data are summarised in Table 1. The first column in this Table provides the values of $E$ and $\eta$ for the simulation. The second and third columns give the best fit values for the configuration and velocity space growth rates, $\Delta r_{rms}$ and $\Delta v_{rms}$, with the errors estimated by fitting the $x$ and $y$ components individually and then averaging. The fourth column yields the average, $\Lambda$, of the configuration and velocity space growth rates. The last column gives the ratio $\Lambda/\chi$, with $\chi$ as computed by Kandrup & Mahon (1994a). There are two entries for $E = 75$ and $\eta = 10^{-6}$ since, for this set of values, two different simulations were performed.

| E | $\eta$ | Position | Velocity | Average | Ratio |
|---|---|---|---|---|---|
| 10 | $10^{-6}$ | 0.204±0.004 | 0.197±0.001 | 0.200±0.004 | 1.504 |
| 30 | $10^{-9}$ | 0.670±0.009 | 0.649±0.001 | 0.659±0.013 | 1.540 |
| 30 | $10^{-6}$ | 0.619±0.031 | 0.582±0.016 | 0.601±0.029 | 1.404 |
| 30 | $10^{-4}$ | 0.629±0.001 | 0.563±0.028 | 0.596±0.041 | 1.392 |
| 75 | $10^{-9}$ | 1.032±0.001 | 1.008±0.002 | 1.020±0.014 | 1.463 |
| 75 | $10^{-6}$ | 0.983±0.009 | 0.924±0.009 | 0.953±0.035 | 1.367 |
| 75 | $10^{-6}$ | 0.977±0.010 | 0.893±0.003 | 0.935±0.049 | 1.341 |
| 75 | $10^{-4}$ | 0.958±0.003 | 0.866±0.002 | 0.913±0.053 | 1.310 |

**Table 1**

Inspection of the data in the Table leads to several obvious conclusions. Overall, as discussed below, the growth rates for both position and velocity are slightly larger for the case of weaker noise. Moreover, the configuration space growth rate tends systematically to be somewhat larger than the velocity space growth rate. However, in every case the ratio $\Lambda/\chi$ is somewhat larger than unity, but decreases slowly with increasing energy. The behaviour of the ratio is easy to understand. The individual initial conditions exponentiate at somewhat different rates and the growth of the sum is dominated by those initial conditions for which the growth rate is above average. One would therefore expect that the ratio should be larger than unity. However, as the energy increases, the relative width of the distribution of local Lyapunov exponents decreases (cf. Kandrup & Mahon 1994b), so that there are fewer initial conditions in which the individual growth rate is much larger than average.

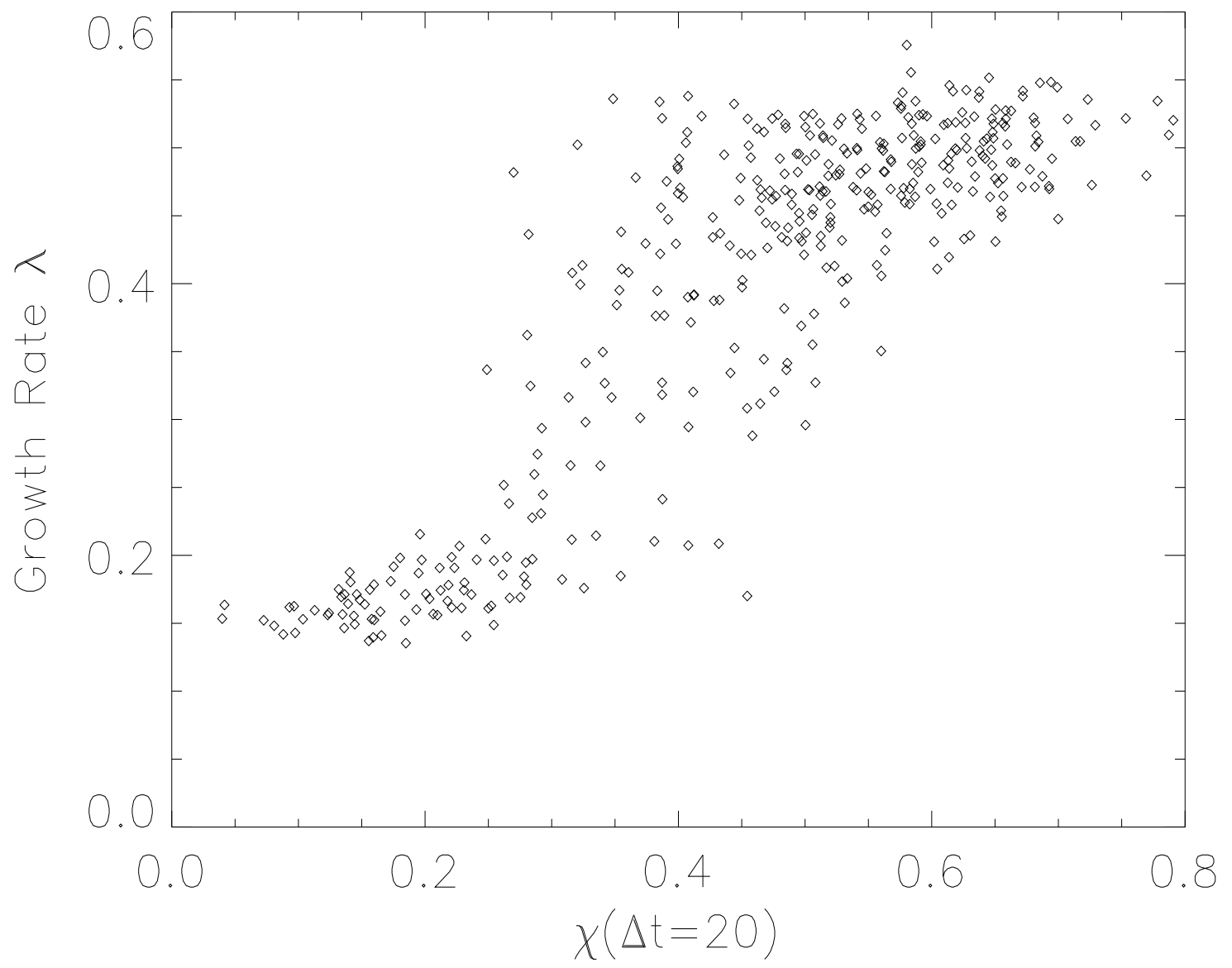

Figure 6: A scatter plot, exhibiting the configuration space growth rate $\lambda_i$ and the local Lyapunov exponent $\chi_i(\Delta t = 20)$ for an ensemble of 400 stochastic initial conditions with $E = 30$, evolved with $\Theta = 30$ and $\eta = 10^{-6}$.

This establishes that there is a direct correlation between the average growth rate $\Lambda$ and the average rate of instability, as given by the Lyapunov exponent $\chi$. However, one might also anticipate that there should be a correlation between the growth rate, $\lambda_i$, for an individual initial condition and the local Lyapunov exponent $\chi_i(\Delta t)$ for that initial condition. One should not expect a complete, one-to-one correlation between these two quantities, since the noise will distort an orbit so as to change the value of the growth rate reflecting the local instability, but one would expect that they should be related. This is confirmed by the data, which show that the correlation is strongest for the case of weak friction and noise. Fig. 6 is a scatter plot of the quantities $\lambda_i$ and $\chi_i(\Delta t = 20)$ for the simulation with $E = \Theta = 30$ and $\eta = 10^{-9}$.

The fact that the correlation is strongest for small $\eta$ is easily understood, given conclusions established in the following section. Larger noise tends to facilitate qualitative changes in the orbits, *e.g.*, triggering large numbers of transitions between different classes of stochastic orbits. However, such qualitative alterations can have the effect of changing the average instability of the orbit.

Combining the results in Sections 3*b* and 3*c*, one infers that, for stochastic orbits, the four moments

$$\delta x_{rms}, \delta y_{rms}, \delta v_{x,rms}, \delta v_{y,rms} \propto (\Theta\eta)^{1/2}\exp(\Lambda t), \tag{13}$$

where $\Lambda$ is comparable in magnitude to the Lyapunov exponent, $\chi(E)$. This is in agreement with the predictions of Pfenniger (1986) and of Kandrup & Willmes (1994).

# 4 The Effects of Friction and Noise on Ensembles of Orbits

## 4.1 Evolution towards a Near-Invariant Distribution

As has been demonstrated elsewhere, both for the truncated Toda potential (Kandrup & Mahon 1994a) and for the $D4$ potential (Mahon *et al* 1994), initially localised ensembles of stochastic orbits with fixed energy $E$ will, when evolved deterministically into the future using the Hamiltonian equations of motion, typically exhibit a coarse-grained exponential evolution towards a near-invariant distribution. Moreover, the characteristic time scale associated with this approach is typically of order ten crossing times $t_{cr}$, which, in "physical" units for a galaxy, corresponds to a time scale much shorter than the age of the Universe. This near-invariant distribution need not correspond exactly to the true invariant distribution associated with the Hamiltonian. However, it is at least approximately invariant in the sense that, although it may exhibit time-dependent oscillations, any subsequent systematic evolution will only proceed on much longer time scales.

The convergence involves a coarse-grained distribution $F(x, y, v_x, v_y, t)$, constructed numerically by binning the orbital data into a four-dimensional grid with fixed cell sizes $\{\Delta x, \Delta y, \Delta v_x, \Delta v_y\}$, and then averaging the output over several successive snapshots. It is particularly convenient to examine the six reduced distributions $f(x, y)$, $f(x, v_x)$, $f(x, v_y)$, $f(y, v_x)$, $f(y,v_y)$, and $f(v_x, v_y)$, generated by summing over the remaining phase space variables. In order to quantify the approach towards an invariant measure, one must specify a measure of "distance" $Df_{1,2}$ between two different distributions, $f_1$ and $f_2$. This is provided by the introduction of a coarse-grained $L^1$ norm.

Here, for example, for two identically normalised distributions, $f_1(x, y)$ and $f_2(x, y)$, binned in an $n \times n$ grid of cells of size $\{\Delta x, \Delta y\}$,

$$Df_{1,2}(x, y) = \frac{\sum_{i=1}^{n} \sum_{j=1}^{n} |f_1(x, y) - f_2(x, y)|}{\sum_{i=1}^{n} f_1(x, y)}. \tag{14}$$

It is with respect to this measure of distance that the Hamiltonian flow evidences an evolution towards a near-invariant measure.

By summing over some large number of snapshots at late times, one can obtain a reasonable sampling of the deterministic near-invariant distribution, $f_{det}$. If the time-dependent $f(t)$ be compared with this $f_{det}$, one observes an approach which is, modulo statistical uncertainties, exponential in time. The rate $\sigma$ associated with this approach appears to be independent of the location and size of the initial ensemble and, at least within a modest range of values, also seems to be independent of the details of the coarse-graining, as well as which of the six possible reduced distributions that one examines. One can thus identify a unique $\sigma(E)$.

On short time scales $\ll \eta^{-1}$, an evolution given by the modified equations (3) involves only minimal changes in the values of the energy and, as such, it is meaningful to speak of a noisy evolution restricted to an "almost constant energy hypersurface." Suppose that one interprets the deterministic evolution towards a near-invariant distribution as a phase mixing associated with the exponential divergence of nearby trajectories. In this case, it is natural to conjecture that an ensemble of noisy orbits generated from a single initial condition will also evolve towards a near-invariant distribution on the same time scale $\sigma$: Small amounts of noise serve as a perturbation that fuzzes out the initial condition into what is effectively a collection of initial conditions, but the subsequent evolution of this ensemble should then be driven by the same exponential instability that is responsible for the deterministic evolution.

This intuition was confirmed for the truncated Toda and $D4$ potentials by analysing large numbers of realisations of individual initial conditions, corresponding to both filling and confined stochastic orbits. For both potentials, it was observed that noisy realisations of a single initial condition, viewed as a single ensemble, do indeed exhibit a coarse-grained evolution towards a near-invariant distribution. If one effects no temporal coarse-graining, simply analysing the data at a sequence of snapshots at intervals $\delta t = 1$, one observes that, at least for the case of weak friction and noise, $\eta \leq 10^{-6}$, the approach towards the invariant measure is not completely monotonic, being characterised instead by damped oscillations. However, if one implements an averaging over several successive oscillations, these oscillations are reduced in amplitude, and one observes a more systematic approach which is better fit by an exponential.

The data also confirm the fact that the rate $\sigma$ associated with this exponential approach is in general independent of the choice of initial condition and, at least roughly, the amplitude of the friction and noise. For the case of initial conditions corresponding to especially sticky stochastic orbits, the rate can be somewhat smaller, at least initially, but, overall, it does make sense to speak of a unique $\sigma(E)$. The best fit value of $\sigma$ also appears to be independent of the details of the coarse-graining. Binning the data into $n \times n$ arrays with different values $n = 10, 20$, and 40 does not induce statistically significant changes in $\sigma$, and different choices of reduced distribution also yield the same value.

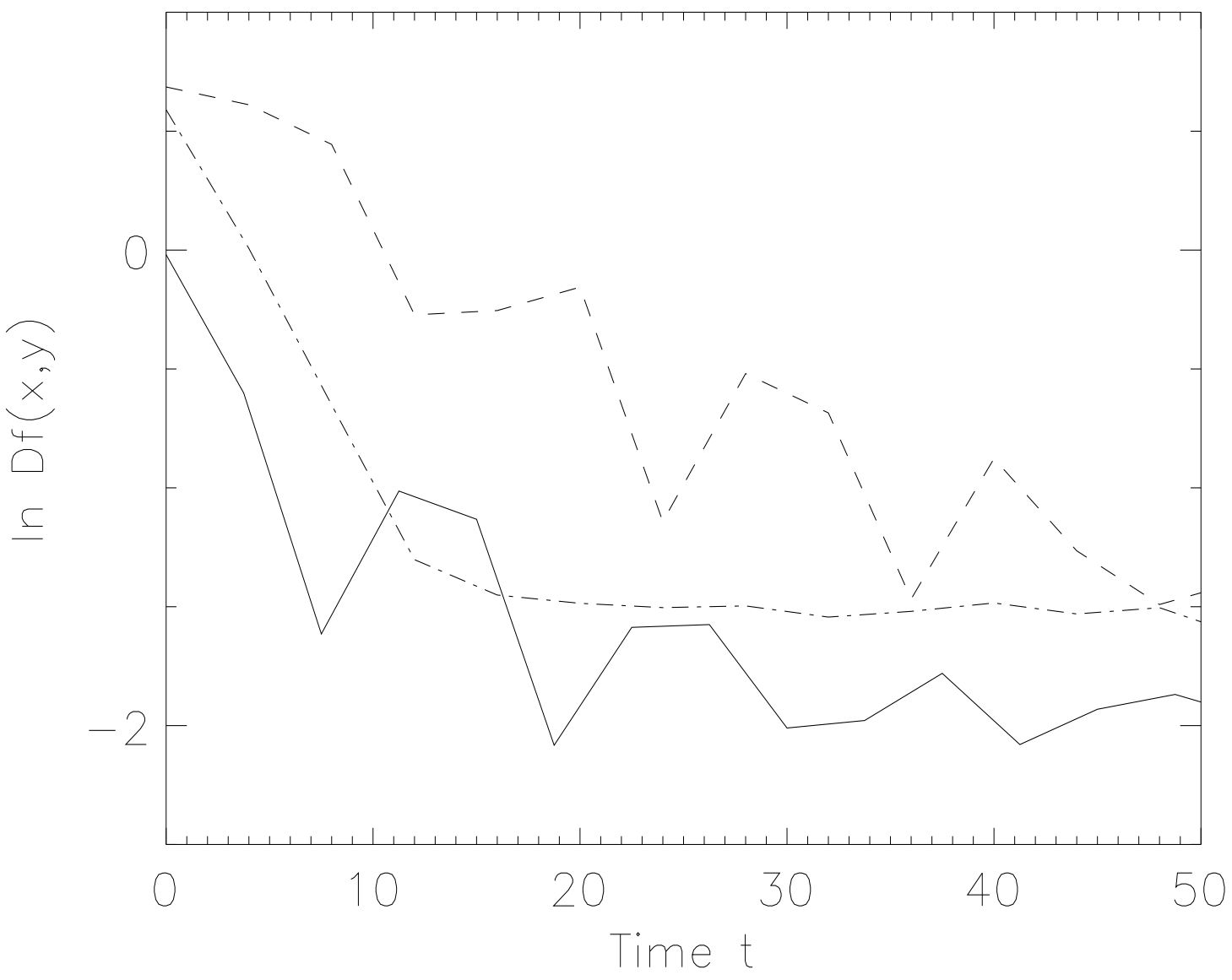

Figure 7: The exponential approach towards a near-invariant distribution. The solid curve represents an unperturbed deterministic evolution with $E = 75$. The dashed and dot-dashed curves represent respectively a noisy simulation with $\eta = 10^{-6}$ and $\eta = 10^{-4}$, in each case with $E = \Theta = 75$. All three curves involve a temporal coarse-graining over intervals $\delta t = 4.0$.

In addition, one observes that the rate associated with this approach towards a noisy near-invariant distribution coincides, to within statistical uncertainties, with the rate associated with the deterministic evolution towards a near-invariant measure. This is, for example, illustrated in Fig. 7, which exhibits data obtained for the truncated Toda potential at $E = 75$. Here the solid curve shows the evolution towards a deterministic near-invariant measure (cf. Kandrup & Mahon 1994a), whereas the remaining curves illustrate the approach towards a noisy near-invariant measure for Langevin simulations with $\Theta = 75$ and variable $\eta = 10^{-6}$ (dashed curve) and $10^{-4}$ (dot-dashed curve). In each of these curves, one sees an initial exponential decrease, which eventually saturates as $Df$ asymptotes towards an approximately constant value. This saturation is a finite size effect. Even if two ensembles sample the same distribution, they will differ because of finite number statistics. When these effects dominate the differences between the two distributions, the exponential convergence must of course terminate. A detailed analysis of the initial exponential decrease, as characterised by the most negative value of the slope, confirms that, in each case, the best fit value of $\sigma$ is approximately the same.

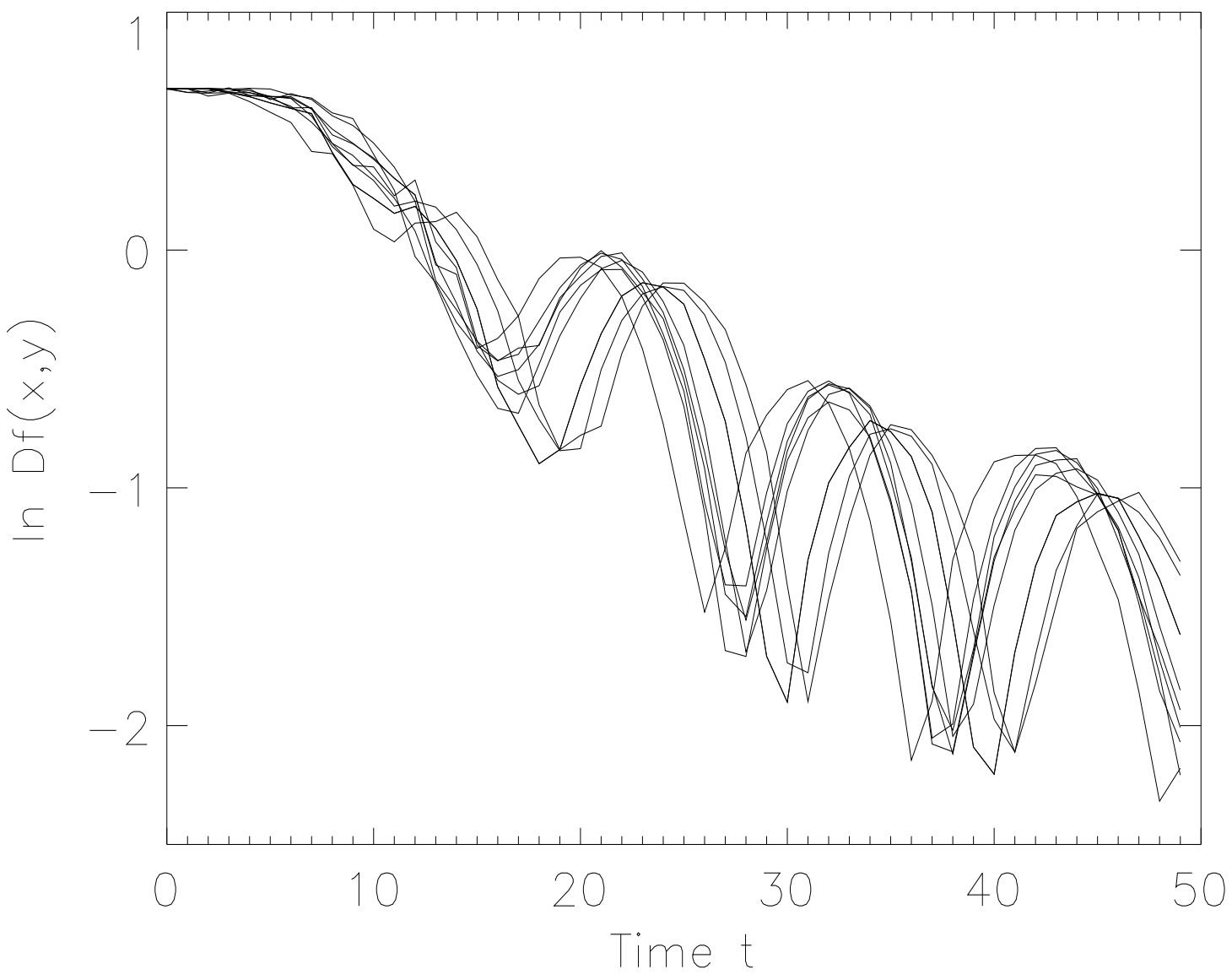

Figure 8: The exponential approach towards a near-invariant measure for nine different initial conditions, evolved with $E = \Theta = 75$ and $\eta = 10^{-6}$. The data were recorded at time intervals $\delta t = 1.0$, but not subjected to any additional coarse-graining.

In Fig. 7, both the deterministic data and the data for the noisy simulations with $\eta = 10^{-6}$ show irregularities which reflect the fact that the evolution towards a near-invariant distribution is not completely monotonic. By contrast, the data for the noisy simulations with $\eta = 10^{-4}$ is approximately monotonic, this a reflection of the fact that, even in the absence of temporal coarse graining, the exponential approach towards the near-invariant measure is more completely nonoscillatory. The oscillatory behaviour associated with raw data not subjected to a time average is exhibited in Fig. 8, which plots $\ln Df(x, y, t)$ for nine different initial conditions with $E = \Theta = 75$, evolved with $\eta = 10^{-6}$. For each individual initial condition, the oscillations are quite conspicuous. However,it is clear that, by effecting either a temporal average or an average over several different initial conditions, these oscillations are systematically suppressed.

The behaviour of regular orbits is very different. Neither for the deterministic evolution of a localised ensemble of initial conditions (cf. Mahon *et al* 1994) nor for multiple noisy realisations of a single regular initial condition does one observe a rapid exponential approach towards a near-invariant distribution. At least for these simple Hamiltonian systems, this rapid approach towards a near-invariant distribution appears to be related directly to the fact that the orbits are stochastic, and thus exhibit an exponential instability towards small changes in initial conditions.

## 4.2 The Form of the Near-Invariant Distribution

Although the deterministic near-invariant measure is, in the sense described above, nearly time-independent, one cannot infer that it necessarily coincides exactly with the true invariant measure, as defined in a $t \to \infty$ limit. Thus, for example, one might suppose that the stochastic portions of the phase space divide into several disjoint regions, and that orbits situated originally in one region will only pass into the other regions very slowly.

There are at least two concrete pieces of numerical evidence to suggest that such a slow secular evolution actually occurs. One of these involves an explicit comparison of the coarse-grained reduced distribution $f(x, y)$ evaluated both at $t = 100$ and at a much later time. For the truncated Toda potential with $E = 30$, samplings of the alleged invariant measure were generated at $t = 100$ and 1100, and the results compared both visually and in terms of the aforementioned $L^1$ norm. The net result (Kandrup & Mahon 1994a) was an indication that the distributions may be slightly different, although the differences are only of marginal statistical significance. An analogous computation was also performed for the $D4$ potential with $E = 6$, comparing $f(x, y)$ at $t = 100$ and 500. In this case, the differences appeared to be somewhat larger.

An easier way to show that the near-invariant distribution is slowly changing is to demonstrate explicitly that some statistical property of the evolving ensemble is varying systematically. Such variations were actually observed for the $D4$ potential at relatively high energies $E \geq 6$. First the near-invariant measure was sampled to generate an ensemble of stochastic initial conditions. These were then evolved deterministically to compute distributions of local Lyapunov exponents for the two intervals $0 \leq t \leq 100$ and $100 \leq t \leq 200$, and the mean values of the distributions extracted. For $E = 6$, the mean value for the first $\Delta t = 100$ was $\overline{\chi} = 0.247 \pm 0.008$, whereas, for the second $\Delta t = 100$, $\overline{\chi} = 0.225 \pm 0.001$. Both these values are considerably larger than the estimated value of the true Lyapunov exponent $\chi = 0.208 \pm 0.010$, obtained by integrating for a total time $t = 10^4$.

These differences can be attributed to the effects of cantori surrounding islands of regularity embedded in the stochastic sea. True KAM tori serve as absolute barriers, separating regular and stochastic orbits. However, cantori are fractured KAM tori, containing a Cantor set of gaps, and thus only impede the passage of orbits, without blocking them completely (cf. MacKay *et al* 1984a,b). If an ensemble of stochastic orbits is originally located entirely outside the cantori, most of the orbits will only breach the gaps on very long time scales, via a so-called intrinsic diffusion (cf. Lichtenberg and Lieberman 1992). This suggests, however, that, on shorter times, one can speak of a near-invariant distribution characterised by a uniform population of the outside regions and a near-zero population of the inside.

The situation is analogous to the classical effusion problem, where two chambers are connected via an extremely narrow conduit. Suppose that an ideal gas is introduced into one of the chambers, and then permitted to evolve. The final equilibrium for the system will correspond presumably to a constant density population of both chambers. However, if the conduit is extremely narrow, the time scale associated with the approach to this equilibrium will be very long, much longer than the time scale on which the gas reaches a near-uniform population in the original chamber. On intermediate time scales, it may therefore make sense to speak of a near-equilibrium, corresponding to a constant density population of the chamber in which the gas was originally inserted and a near-zero density population of the remaining chamber.

What makes this phenomenon important in the present context is the possibility that friction and noise can dramatically accelerate this process, augmenting the aforementioned intrinsic diffusion by an additional extrinsic diffusion (cf. Lieberman and Lichtenberg 1972). This possibility, which has obvious implications for the form of the near-invariant measure, was examined in detail for both the truncated Toda potential and the $D4$ potential.

For energies where the phase space is mostly stochastic, and where the islands of regularity are small, the deterministic and noisy near-invariant measures appear to be very similar. However, for energies where the relative measure of regular orbits is larger, so that one might expect that cantori are more important, the noisy and deterministic near-invariant measures are distinctly different. The most dramatic examples obtain for the $D4$ potential with $E \geq 6$.

The case $E = 15$ is illustrated in Figs. $9a - c$. The first panel shows a grey scale plot of the deterministic near-invariant measure, using a linear scale for intensity. The second shows an identically normalised plot of the noisy near-invariant measure, generated with $\Theta = 15$ and $\eta = 10^{-6}$. The difference between the two distributions is exhibited in the final panel. It is clear by visual inspection that the two distributions exhibit systematic differences. These can be attributed to sticky stochastic orbits which have become trapped around two families of regular orbits, one of which passes near the origin along an approximately figure eight trajectory and the other a rosette-shaped family that avoids the central regions. The fact that the distributions are different can also be quantified by computing the $L^1$ distance between them, and verifying that this is larger than can be attributed to statistical effects.

A comparison of the data obtained from deterministic and noisy simulations indicates that friction and noise lead to an evolution towards a noisy near-invariant measure which is closer to the deterministic true invariant measure than to the deterministic near-invariant measure realised on shorter times $t \leq 100$. Unlike the short time deterministic near-invariant measure, which is dominated by a single population of filling stochastic orbits, both the other measures seem comprised of two distinct populations, one corresponding to filling stochastic orbits and the other to confined stochastic orbits. Moreover, at least for the case of the $D4$ potential, there is unambiguous evidence that the two different populations are characterised by markedly different distributions of local Lyapunov exponents, the typical $\chi$ for the confined orbits being substantially smaller in value than the $\chi$'s associated with filling orbits.

To illustrate this latter point, distributions of local Lyapunov exponents were computed in three different ways. (1) The deterministic near-invariant measure was sampled to generate an ensemble of initial conditions, and these were then evolved to compute a distribution $N_{det}(\chi(\Delta t{=}100))$. (2) The noisy near-invariant measure was used to compute an analogous distribution $N_{noisy}(\chi(\Delta t{=}100))$. (3) Several long time integrations for a total time $t = 10^4$ were partitioned into segments of length $\Delta t = 100$ to extract short time exponents $\chi(\Delta t = 100)$, and these then combined to extract a long time distribution $N_{long}(\chi(\Delta t{=}100))$.

Typically, the distribution $N_{det}$ is unimodal and roughly Gaussian in shape, although, particularly for energies containing a large measure of regular orbits, it can acquire a low $\chi$ tail (cf. Kandrup & Mahon 1994, Mahon *et al* 1994). When the regular regions are relatively small, the distributions $N_{noisy}$ and $N_{long}$ closely resemble $N_{det}$. However, when the regular regions become larger, and cantori become important, these distributions become significantly different. The long time integration leads typically to a distribution $N_{long}$ which contains a larger proportion of low $\chi$ contributions and, for sufficiently large regular regions, becomes distinctly bimodal, with one peak corresponding to the $\chi$ at which $N_{det}$ is maximum and another peak at a much small $\chi$. In this case, the noisy distribution $N_{noisy}$ looks very different from $N_{det}$, tending overall to resemble $N_{long}$ much more closely.

This behaviour is illustrated in Figs. 10 and 11, which exhibit the three different distributions for the $D4$ potential with, respectively, $E = 6$ and 15. It is clear by inspection that, for $E = 6$, the distributions $N_{long}$ and $N_{noisy}$ look relatively similar to one another, characterised by bimodal distributions with maxima at the same values of $\chi$. The unimodal distribution $N_{det}$ is quite different. In this case, $N_{det}$ is characterised by a mean $\overline{\chi} = 0.247 \pm 0.008$, $N_{long}$ by $\overline{\chi} = 0.208 \pm 0.010$, and $N_{noisy}$ by $\overline{\chi} = 0.182 \pm 0.007$.

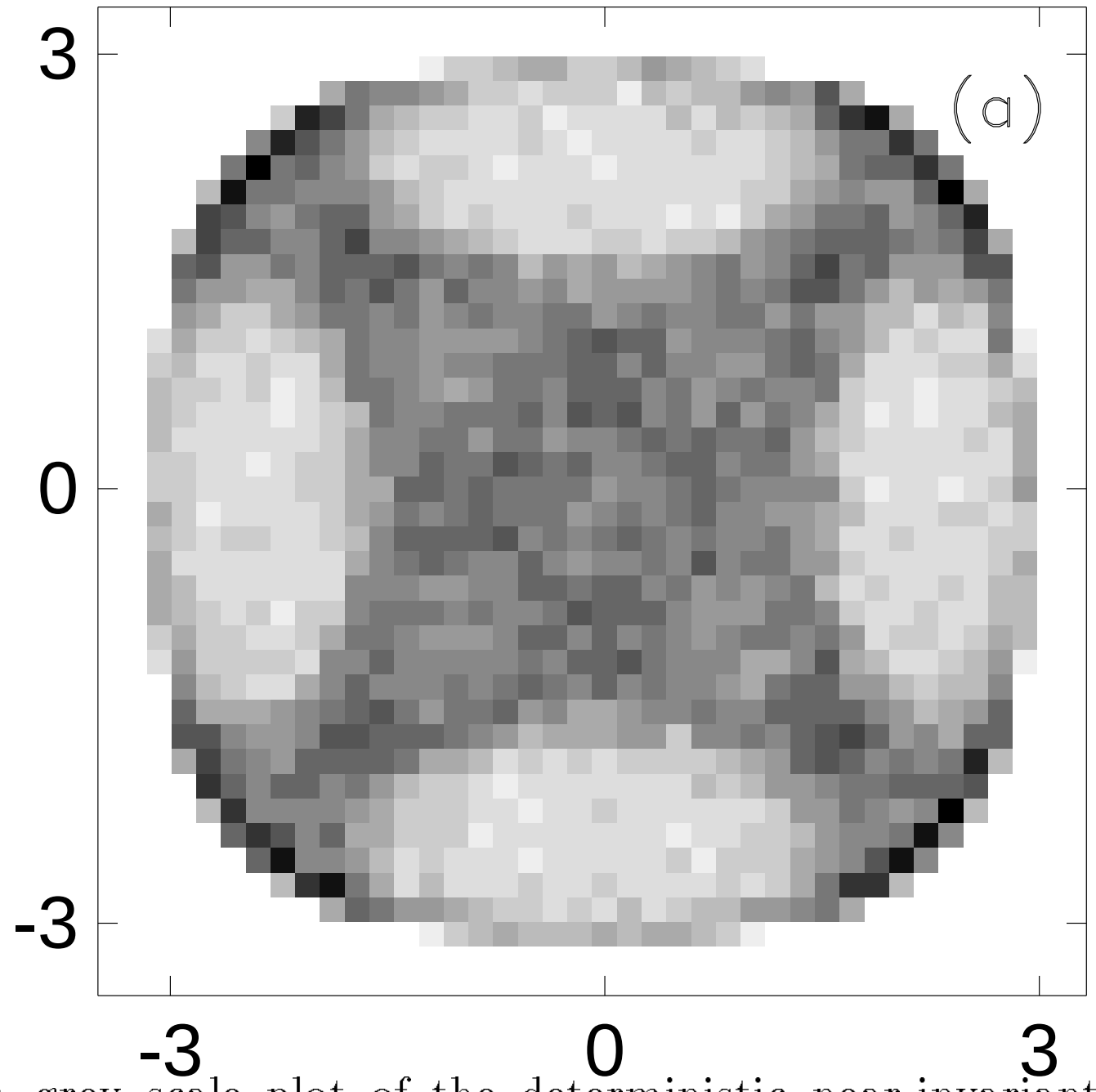

Figure 9: (a) A grey scale plot of the deterministic near-invariant $f_d(x,y)$ for the $D4$ potential with $E = 15$, generated from a $40 \times 40$ binning. Darker shades represent higher densities.

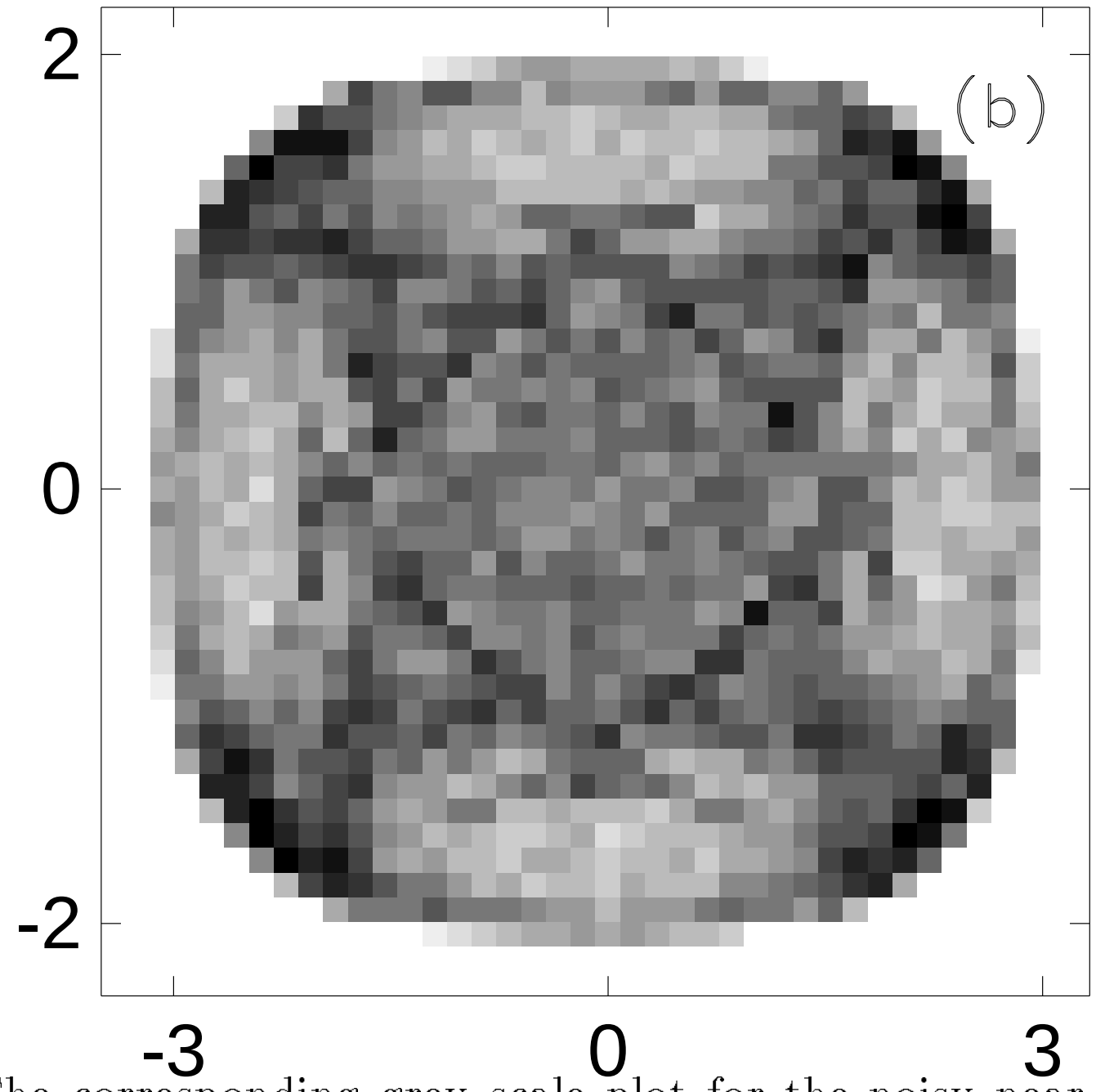

Figure 9: (b) The corresponding grey scale plot for the noisy near-invariant $f_\eta(x,y)$ for $E = \Theta = 15$ and $\eta = 10^{-6}$.

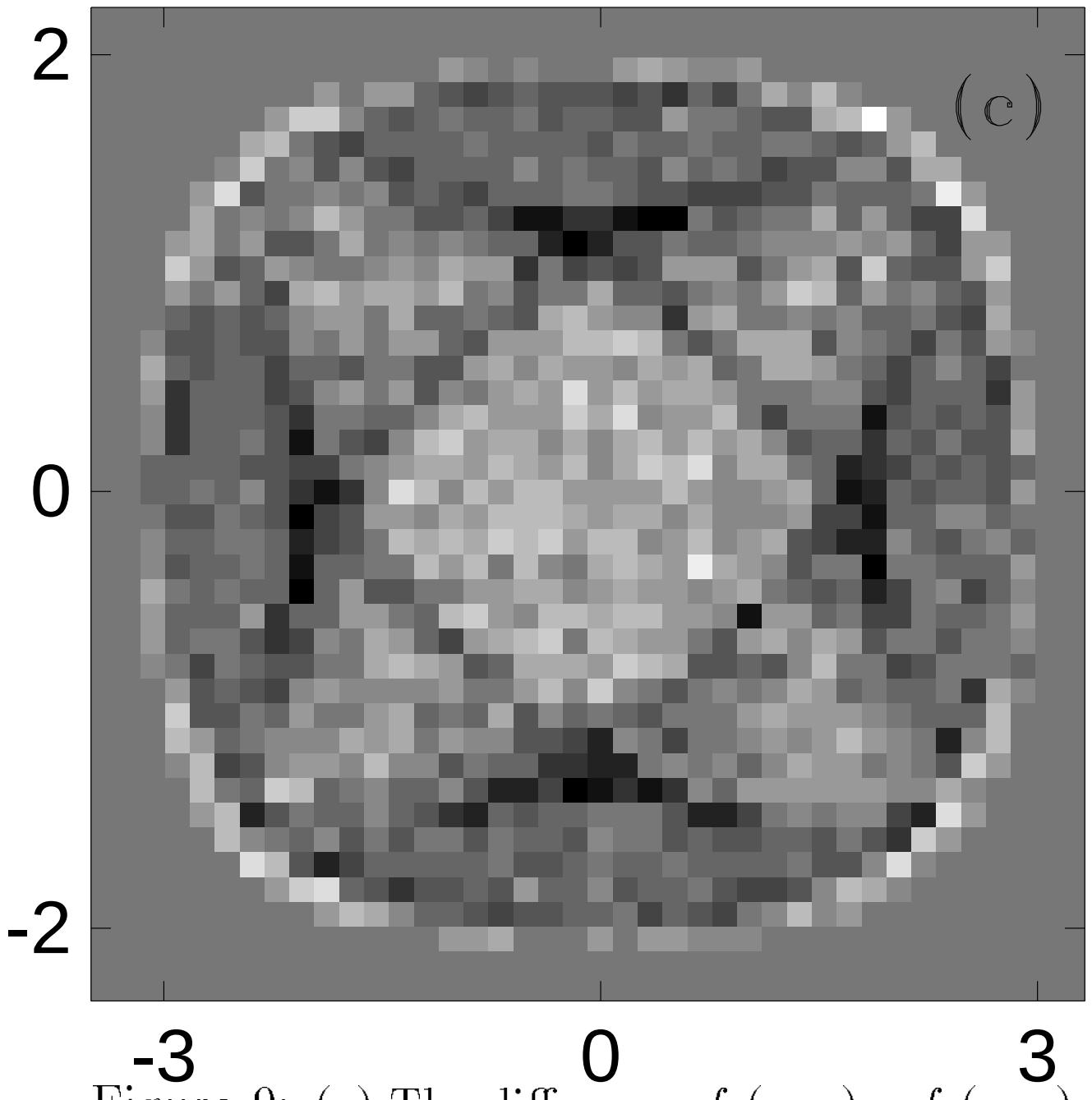

Figure 9: (c) The difference $f_d(x, y) - f_\eta(x, y)$.

For $E = 15$, $N_{long}$ and $N_{noisy}$ seem at first glance to be substantially different from one another. However, they are similar to each other, and different from $N_{det}$, in that they both have a bimodal population, although, for $N_{long}$ the lower component has a substantially wider dispersion. Moreover, despite this obvious difference, these two distributions are similar in that they are characterised by very similar means. In this case, $N_{det}$ has a mean $\overline{\chi} = 0.243 \pm 0.004$, $N_{long}$ has $\overline{\chi} = 0.158 \pm 0.018$, and $N_{noisy}$ has $\overline{\chi} = 0.155 \pm 0.001$.

For both $E = 6$ and $E = 15$, especially the latter case, it is possible that the statistics are contaminated to a certain extent by the fact that the noisy initial conditions used to generate the solid curves actually contain a few regular orbits. Although transitions from stochastic to regular are typically uncommon for $\eta \ll 10^{-4}$, they can occur; and they will be less uncommon for energies where the stochastic phase space regions are substantially smaller than the regular regions.

The experiments summarised above involved multiple realisations of a single initial condition. However, one can also sample the deterministic near-invariant distribution, and then evolve it into the future, allowing for the effects of friction and noise. This was done for both the truncated Toda and $D4$ potentials. Not suprisingly, one observes a systematic evolution away from the deterministic near-invariant measure $N_{det}$ and towards the noisy near-invariant distribution $N_{noisy}$. Moreover, as reported in Habib *et al* (1994a), one observes that the evolution from one near-invariant measure to the other proceeds more rapidly when the friction and noise are larger in amplitude. Consider, for example, an ensemble with initial energy $E = 30$, evolving in the truncated Toda potential. Here, for simulations with $\Theta = 30$ and $\eta = 10^{-6}$ and $10^{-9}$, there is clear evidence of an evolution towards the noisy near-invariant measure within a time $\sim 100 - 200$. For $\eta = 10^{-4}$, changes in energy induced by the perturbations are sufficiently large that one can no longer speak of an approximately constant energy hypersurface. However, the data are still consistent with an evolution in which the cantori surrounding the regular islands have been breached.

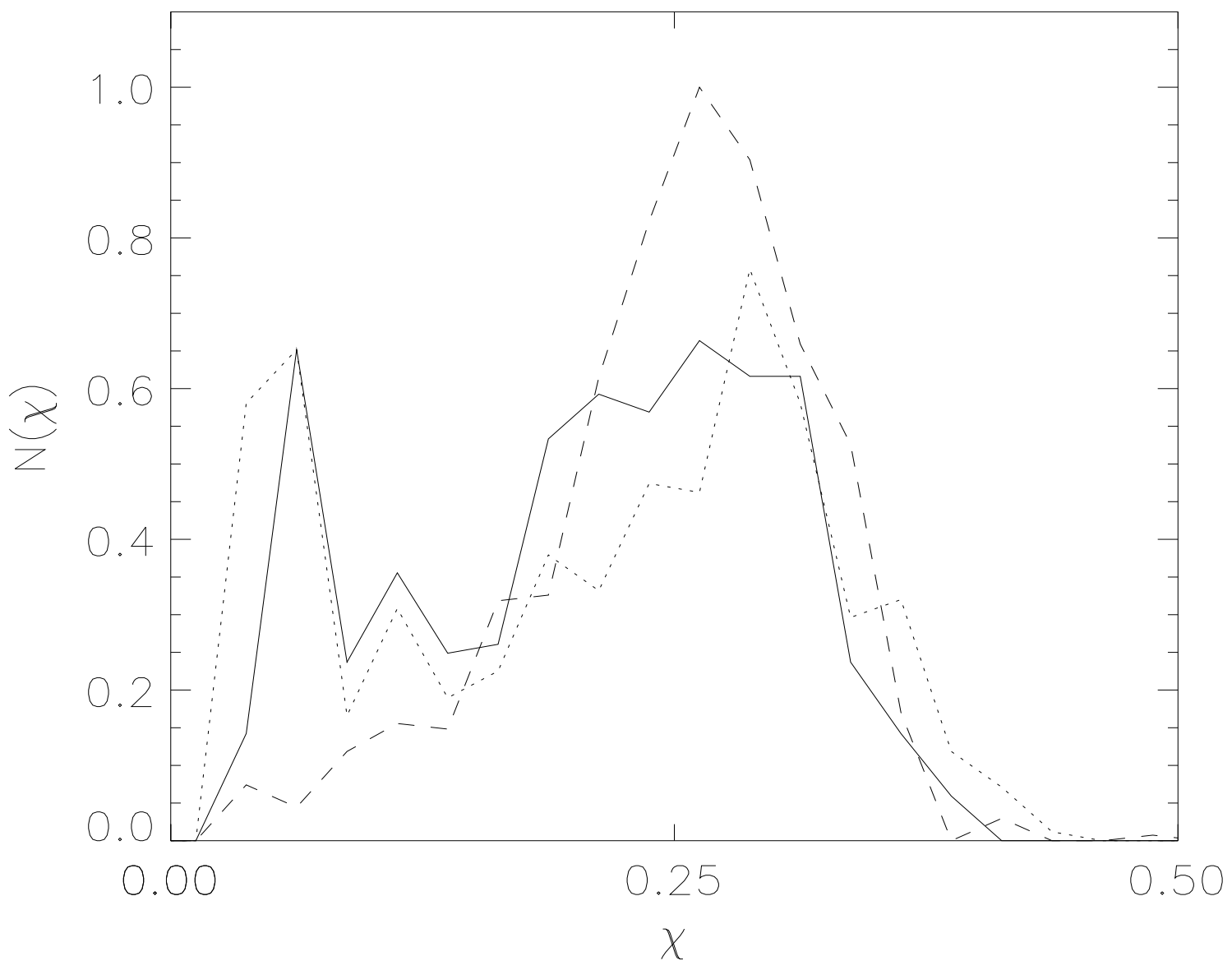

Figure 10: The dashed curve gives the binned distribution of local Lyapunov exponents generated from a sampling of the deterministic near-invariant measure for the $D4$ potential with $E = 6$. The solid curve gives the corresponding distribution for the noisy near-invariant measure generated with $E = \Theta = 6$ and $\eta = 10^{-6}$. The dotted curve gives the distribution of local Lyapunov exponents generated by partitioning several long time ($t = 10000$) deterministic computations.

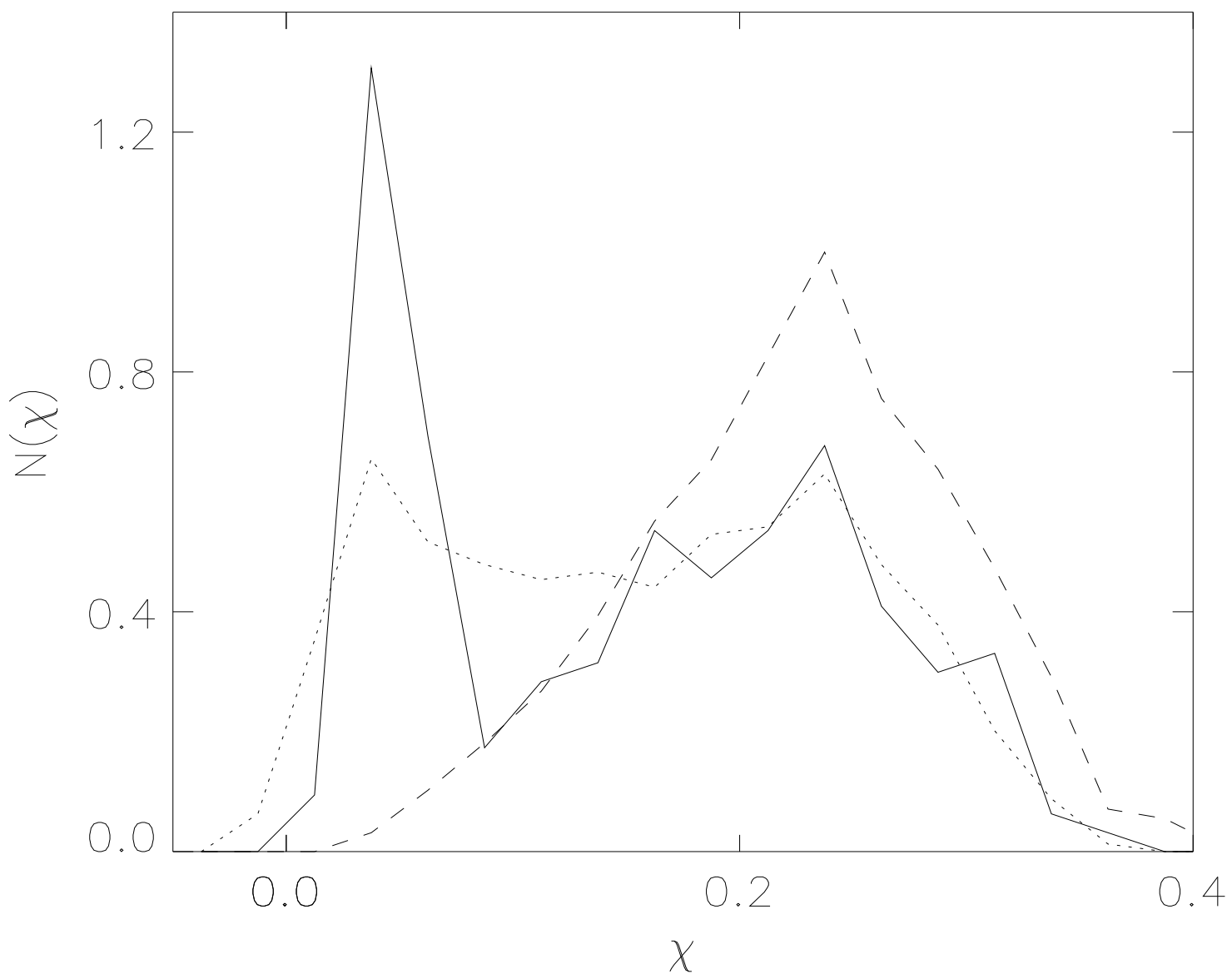

Figure 11: The analogue of Fig. 10, generated for $E = \Theta = 15$ and $\eta = 10^{-6}$

Both for the truncated Toda and the $D4$ potentials, analysis of the data also indicates that the evolution from $N_{det}$ to $N_{noisy}$ is approximately exponential. For example, for the truncated Toda potential with $E = 30$, using a $10 \times 10$ binning of $f(x, y)$, the best fit values for the slope are: $\mu = 0.0143$ for $\eta = 10^{-9}$, $\mu = 0.0339$ for $\eta = 10^{-6}$, and $\mu = 0.0460$ for $\eta = 10^{-4}$.

## 4.3 Changes in Orbit Class

The conclusions articulated hitherto imply that even relatively weak friction and noise are sufficient to induce large numbers of transitions between filling and confined stochastic orbits within a time $t < 100$. By contrast, similar transitions are not observed between regular and stochastic orbits. Only for $\eta$ as large as $\sim 10^{-3}$ does one see frequent transitions between regular and stochastic orbits on these short time scales.

It is difficult to construct a simple, automatable algorithm to decide precisely when a stochastic orbit has changed from filling to confined, or vice versa; and, for this reason, it is difficult to determine a precise numerical value for the "average transition rate." However, visual inspection of a very large number of orbits ($\sim 2.5 \times 10^4$) in both the truncated Toda and $D4$ potentials leads to simple, semi-quantitative conclusions which corroborate the preliminary results reported by Kandrup & Mahon (1994c).

For those energies where the accessible phase space is almost completely stochastic, the sticky regions, if they exist at all, are only very small in volume. One therefore observes relatively few transitions from filling to confined orbits, even if the deterministic evolution is perturbed by relatively large friction and noise. However, at energies where the regular regions are larger, the relative measure of sticky stochastic orbits is also larger, and, for these energies, even very weak friction and noise can have significant effects.

To characterise the overall importance of these effects on relatively short time scales, one can, for example, estimate the fraction of the orbits which are altered qualitatively by friction and noise of given amplitude within a time $t = 100$. Consider, for specificity, energies $E \sim 20$ in the truncated Toda potential or $E \sim 10$ in the $D4$ potential, which lead to comparable results. Suppose further that $\Theta \sim E$. In this case, one finds that, for $\eta$ as small as $10^{-12}$, the friction and noise have no appreciable effects in either potential. However, when $\eta$ is increased to a value $\sim 10^{-9}$, one begins to observe some transitions and, even for $\eta$ as small as $10^{-6}$, the effects of noise are quite conspicuous. Indeed, for $\eta \sim 10^{-6}$, more than half the orbits exhibit some sort of "obvious" visual signal emblematic of a significant qualitative change: Confined orbits can become filling, filling orbits can become trapped around some family of regular orbits, and one even sees confined orbits originally trapped around one family of regular orbits become trapped instead around a different family of regular orbits.Some representative examples of noisy trajectories in the truncated Toda potential with $E = \Theta = 20$ and $\eta = 10^{-6}$ are exhibited in Kandrup & Mahon (1994c). If $\eta$ is increased to yet larger values, transitions become even more abundant; and, for $\eta$ as large as $10^{-3}$, transitions are so frequent that the distinction between filling and confined seems almost meaningless.

As stated earlier, when translated into "physical" units appropriate for a galaxy, the choice of $\eta = 10^{-6}$ is extremely suggestive. The characteristic crossing time for a galaxy is $t_{cr} \sim 10^8$yr, so that the age of the Universe $t_H \sim 100 t_{cr}$. A relaxation time $t_R = 10^{14}$yr thus corresponds to $10^6 t_{cr}$, so that $\eta = t_{cr}/t_R \sim 10^{-6}$. For noise associated with external influences, the effective relaxation time could be much smaller.

# 5 Conclusions and Potential Implications

As discussed in Section 1, realistic perturbations of orbits in a smooth potential can oftentimes be modeled as friction and noise, related via a fluctuation-dissipation theorem. It is therefore natural to investigate the effects – both quantitative and qualitative – of friction and noise on the form of these orbits. As discussed in Section 2, at present there is no compelling evidence to exclude and there *are* indications that such global stochasticity may exist. It is therefore important to explore the implications of the hypothesis that galactic potentials admit both regular and stochastic orbits.

The aim of this paper has been to investigate the effects of friction and noise on orbits in two different, strongly nonintegrable two-dimensional potentials, which admit appreciable numbers of both regular and stochastic orbits. These potentials are very different from each other. The fact that they exhibit similar qualitative behaviour therefore suggests that the principal conclusions derived here should be generic for nonintegrable potentials. In particular, one might expect that the conclusions rely only on such general topological properties as the existence of KAM tori and cantori.

Viewed in energy space or, presumably, in terms of the other collisionless invariants, friction and noise serve as a classical diffusion process, acting on all orbit classes in an identical fashion and only inducing appreciable changes in the collisionless invariants on the natural time scale $t_R \sim \eta^{-1}$. However, when viewed in configuration or velocity space, the effects of friction and noise are more complicated, and depend on the class of orbit. In stochastic orbits, friction and noise trigger instabilities which grow exponentially at a rate that is given by the local Lyapunov exponent. By contrast, in regular orbits friction and noise serve as perturbations that only grow as a power law in time, albeit with an exponent that is larger than the value predicted for the integrable harmonic potential. However, despite these differences, there is one common feature, namely the fact that the second moments, $\delta x_{rms}$, $\delta y_{rms}$, $\delta v_{x,rms}$, and $\delta v_{y,rms}$ all scale as $(\Theta\eta)^{1/2}$.

The exponential instability triggered by friction and noise implies that an ensemble of noisy stochastic orbits, generated from the same initial condition, will exhibit a coarse-grained exponential evolution towards a noisy near-invariant distribution at the same rate at which initially localised ensembles of stochastic initial conditions evolve deterministically towards a near-invariant distribution. However, the noisy and deterministic near-invariant distributions can be significantly different in form. Thus, if one samples the deterministic near-invariant distribution to generate a set of initial conditions, and evolves these initial data into the future allowing for friction and noise, one can observe statistically significant changes already on very short time scales $\ll t_R$. These changes reflect the fact that even very weak friction and noise, corresponding to a relaxation time $t_R \gg t_H$, can trigger significant numbers of changes in orbit class within a time $t_H$, thereby dramatically accelerating the rate at which trajectories diffuse through cantori.

These rigourous results may have concrete implications for problems in galactic dynamics. One point is that even relatively weak friction and noise may accelerate the approach towards a true self-consistent equilibrium, reducing the time required from $t \gg t_H$ to a time $t \ll t_H$. The calculations in this paper are not self-consistent, but they *do* suggest that small, non-Hamiltonian irregularities can accelerate various evolutionary processes by serving as a source of extrinsic diffusion.

More generally, it is clear that the existence of chaotic orbits can give rise to qualitatively new behaviour. The phase space associated with a strongly nonintegrable potential is substantially more complicated than the phase space associated with an integrable or near-integrable potential, being partitioned by both absolute and partial barriers. In the absence of friction and noise the partial barriers may be almost absolute on short time scales $\ll t_H$, but friction and noise can substantially accelerate the time scale on which these partial barriers are traversed.

The exponential divergence observed for stochastic orbits quickly renders a deterministic integration in a smooth potential meaningless, both pointwise and, in certain cases, statistically. This fact has direct and potentially significant implications for the problem of shadowing. Recently, astronomers have become concerned about the question of whether, and in what sense, numerical solutions to the $N$-body problem, or any other set of differential equations, accurately reflect the true solution to the equations that one is trying to solve (cf. Quinlan & Tremaine 1993). However, the simulations described here suggest another, equally important question of shadowing (cf. Eubank & Farmer 1990). To what extent can the "real world," which is characterised by complicated irregularities, shadow either one's model system or numerical realisations thereof? The answer to this is unclear. However, the simulations summarised here might suggest that, rather than being a major impediment to realistic modelling, numerical noise may not always be a bad thing, in that it may capture, at least qualitatively, some of the effects of the small perturbing influences to which real systems are always subjected.

Conventional wisdom suggests that a useful "indication" of stability or instability for a collisionless equilibrium comprised of regular orbits is whether or not regular orbits in the equilibrium potential are stable. Such a test is of course meaningless for individual stochastic orbits since they are all unstable. However, it does make sense to ask whether an *ensemble of stochastic orbits* is stable towards small external perturbations. The simulations in this paper would suggest that, oftentimes, the answer to this may be: no.

When studying the evolution of orbits in a fixed time-independent potential, possibly perturbed by friction and noise, one can identify four different notions of invariant, or near-invariant, measure, viz: (1) the true deterministic invariant measure, (2) a deterministic near-invariant measure, (3) the true noisy invariant measure, and (4) a noisy near-invariant measure. The true noisy invariant measure is of course irrelevant on short time scales since, as discussed already, friction and noise only induce significant changes in the collisionless invariants on times $\sim t_R$.

In the context of galactic dynamics, it might seem natural in the first instance to focus on the true deterministic invariant measure, since this corresponds to a true equilibrium for the Hamiltonian evolution. However, there is no guarantee that, on short time scales, this distribution will be achieved. Viewed over sufficiently long time scales, there are only two different classes of orbits, namely regular orbits and stochastic orbits. However, on shorter time scales, deterministic stochastic orbits divide naturally into two, essentially distinct, subclasses, confined stochastic orbits and unconfined stochastic orbits; and, because these subclasses are distinct, it makes sense to speak of separate confined and unconfined near-invariant measures. It might therefore seem more appropriate to consider instead the possibility of a quasi-equilibrium, characterised by a deterministic near-invariant measure. Perhaps the most important conclusion derived from the simulations summarised in this paper is that the noisy near-invariant measure may be the more appropriate choice.

Theoretically there is reason to suppose that the noisy near-invariant measure will closely approximate the true deterministic invariant measure, *i.e.*, that the only real effect of friction and noise is to accelerate the rate at which orbits breach cantori by providing a source of extrinsic diffusion. On short time scales $\ll t_R$, a noisy evolution is still restricted, at least approximately, to a lower-dimensional hypersurface, determined by the value of the energy and any other collisionless invariants. It follows that the only possible effect of the friction and noise can be to readjust the population of orbits in this hypersurface. However, to the extent that the evolution entails a coarse-grained approach towards a uniform population of the accessible regions, the only effects that the friction and noise can have should be to change which regions are accessible on short time scales.

When considering the effects of friction and noise on ensembles of orbits diffusing through cantori, there are three potential sources of asymmetry, namely 1) the properties of the cantori, 2) the properties of the friction and noise, and 3) the choice of initial data. The standard "turnstile model" of cantori (cf. MacKay *et al* 1984a,b), which has met with considerable success in explaining the results of numerical experiments, introduces no asymmetries in the cantori *per se*. Passing through a cantorus should be just as likely from one direction as from the other. Friction and noise can in principle serve as a source of asymmetry. For example, one could suppose that the friction and noise are so chosen so as to vanish identically for orbits on one side of the cantorus, but to be nonvanishing and large for orbits on the other. However, this seems contrived. To the extent that the friction and noise vary smoothly as a function of phase space coordinates – as for the case of additive noise and a constant coefficient of dynamical friction – the influences acting on an orbit just inside and just outside a cantorus should be essentially identical.

It follows that the only source of asymmetry in the simulations described here is provided by the choice of initial conditions. If an ensemble of orbits is initially located inside a cantorus, individual members will systematically escape; if the ensemble of orbits is initially located outside, individual orbits will tend instead to be trapped systematically. In either case, however, the detailed memory of initial conditions will eventually be erased, as the ensemble evolves towards a final quasi-equilibrium which includes both confined and filling stochastic orbits. As discussed, *e.g.*, in Section 4, the numerical simulations corroborate this simple argument, suggesting that, in fact, the noisy near-invariant distribution is quite similar to the true deterministic invariant distribution. In this sense, it would appear that, on short time scales, the principal effect of friction and noise is to accelerate the evolution towards the true deterministic invariant distribution.

In the past, various authors (cf. Contopoulos 1993, Kaufmann 1993) have sought to use confined stochastic orbits as one of the building blocks for the construction of approximate self-consistent equilibria, the idea being that such orbits can play the role of regular orbits in regions near corotation and other resonances where no true regular orbits exist. However, the calculations presented herein suggest that models incorporating such confined stochastic orbits may be unstable towards an outwards extrinsic diffusion, whereby the confined orbits escape to become filling. If a model incorporates both filling and stochastic orbits with appropriately chosen relatively abundances, it may be stable towards noise-induced transitions between filling and confined. However, such stability hinges on the existence of a detailed balance for transitions between these different orbit classes.

Suppose, for example, that an approximate self-consistent equilibrium contains a bar supported in part by confined stochastic orbits (cf. Wozniak 1993). Even neglecting discreteness effects and all other perturbations, the confined stochastic orbits will eventually leak out, although the characteristic time scale for this could be $\gg t_H$. One might therefore speak of a bar which, albeit formally unstable, is effectively stable on astrophysically relevant time scales $< t_H$. However, if one allows for even weak friction and noise, the leakage time scale can be much shorter. Significant numbers of orbits may escape within a time $< t_H$, so that crucial support for the bar is lost and the structure eventually dissolves. Alternatively, as suggested by Fig. 9, if a model also contains large numbers of filling stochastic orbits, friction and noise could induce trapping in the neighbourhood of the regular orbits, thereby causing the bar to become yet more pronounced.

One final caveat should be stressed. All the simulations reported herein were effected for two-dimensional, rather than three-dimensional, potentials. However, one knows that Hamiltonian systems with dimensionality $D > 2$ are different from two-dimensional systems in that they admit Arnold diffusion (cf. Lichtenberg &

Lieberman 1992). Nevertheless, the characteristic time scale associated with Arnold diffusion can be very long (cf. Nekhoroshev 1977), so that, in the absence of friction and noise, orbits in three-dimensional systems will oftentimes exhibit the same qualitative behaviour as orbits in two-dimensional systems. Whether friction and noise can also accelerate processes like Arnold diffusion is an interesting unsolved question, although there are indications (cf. Tennyson 1982) that the answer may oftentimes be: yes.

## 6 Acknowledgments

The authors acknowledge useful discussions with Robert Abernathy, Paul Channell, Edward Ott, and Daniel Pfenniger. SH was supported in part by the DOE and by AFOSR. HEK was supported in part by the NSF grant PHY92-03333. MEM was supported by the University of Florida. Computer time was made available through the Research Computing Initiative at the Northeast Regional Data Center (Florida) by the IBM Corp. and on the CM-5, by the Advanced Computing Laboratory at Los Alamos National Laboratory.

# A Appendix

The Langevin equations (3) are equivalent (cf. Chandrasekhar 1943c, van Kampen 1981) to a Fokker-Planck equation

$$\frac{\partial f}{\partial t} + \mathbf{v}\cdot\frac{\partial f}{\partial \mathbf{r}} - \nabla\Phi\cdot\frac{\partial f}{\partial \mathbf{v}} = -\frac{\partial}{\partial \mathbf{v}}\cdot\left(\eta\mathbf{v}f\right) + 2\Theta\,\frac{\partial^2 f}{\partial \mathbf{v}^2}. \qquad (A1)$$

Given eq. (A1) it is easy to derive moment equations of the form

$$\frac{d\langle X\rangle}{dt} = .... \qquad (A2)$$

for various physical quantities $X = x^{p_1} y^{p_2} v_x^{p_3} v_y^{p_4}$. In particular, it is easy to see that, for a harmonic oscillator potential, $\Phi = \frac{1}{2}\mathbf{r}^2$,

$$\frac{d\langle E\rangle}{dt} = \eta\Big(\Theta - \langle \mathbf{v}^2\rangle\Big) \tag{A3}$$

and

$$\frac{d\langle E^2\rangle}{dt} = -\eta\Big(\langle \mathbf{v}^4\rangle + \langle \mathbf{r}^2\mathbf{v}^2\rangle\Big) + \eta\Theta\Big(3\langle \mathbf{v}^2\rangle + \langle \mathbf{r}^2\rangle\Big), \tag{A4}$$

so that

$$\frac{d\langle(\delta E)^2\rangle}{dt} = 2\eta\Theta\langle \mathbf{v}^2\rangle + \eta\Big(\langle \mathbf{v}^2\rangle^2 - \langle \mathbf{v}^4\rangle\Big) + \eta\Big(\langle \mathbf{v}^2\rangle\langle \mathbf{r}^2\rangle - \langle \mathbf{v}^2\mathbf{r}^2\rangle\Big). \tag{A5}$$

Suppose now that one is concerned with early time behavior in the limit of weak friction and noise. At sufficiently early times, the moments of the noisy trajectories closely track the deterministic trajectory, so that, *e.g.*, $\langle \mathbf{v}^4\rangle \approx \langle \mathbf{v}^2\rangle^2 \approx \mathbf{v}_0^4$. Moreover, given sufficiently weak friction and noise, one can average over effects proceeding on a typical orbital period to infer that $\frac{1}{2}\langle \mathbf{v}^2\rangle \approx \frac{1}{2}\langle \mathbf{r}^2\rangle \approx \frac{1}{2}\langle E\rangle$, which, at early times, reduces to $\frac{1}{2}E_0$. It follows that

$$\frac{d\langle(\delta E)^2\rangle}{dt} \approx 2\eta\Theta E_0, \tag{A6}$$

from which eq. (7) is immediate.